\documentclass[preprint,aps,pra,showkeys,unsortedaddress,superscriptaddress,longbibliography]{revtex4-1}
\usepackage{subfiles}
\usepackage{graphicx}
\usepackage{amsfonts,amsmath}
\usepackage{geometry}
\usepackage[table]{xcolor}

\AtBeginDocument{%
    \newwrite\bibnotes
    \def\bibnotesext{Notes.bib}
    \immediate\openout\bibnotes=\jobname\bibnotesext
    \immediate\write\bibnotes{@CONTROL{REVTEX41Control}}
    \immediate\write\bibnotes{@CONTROL{%
    apsrev41Control,author="08",editor="1",pages="1",title="0",year="1"}}
     \if@filesw
     \immediate\write\@auxout{\string\citation{apsrev41Control}}%
    \fi
}%

\begin{document}
\title{Quasi-steady-state measurement of exciton diffusion lengths in organic semiconductors }
\author{Drew B. Riley}
\email[Email Drew B. Riley at: ]{1915821@swansea.ac.uk}
\author{Oskar J. Sandberg}
\email[Email Dr. Oskar J. Sandberg at: ]{o.j.sandberg@swansea.ac.uk}
\affiliation{Sustainable Advanced Materials Programme (Sêr SAM), Department of Physics, Swansea University, Singleton Park, Swansea SA2 8PP, United Kingdom}
\author{Wei Li}
\author{Paul Meredith}
\affiliation{Sustainable Advanced Materials Programme (Sêr SAM), Department of Physics, Swansea University, Singleton Park, Swansea SA2 8PP, United Kingdom}
\author{Ardalan Armin}
\email[Email Dr. Ardalan Armin at: ]{ardalan.armin@swansea.ac.uk}
\affiliation{Sustainable Advanced Materials Programme (Sêr SAM), Department of Physics, Swansea University, Singleton Park, Swansea SA2 8PP, United Kingdom}

\begin{abstract}
Exciton diffusion plays a decisive role in various organic optoelectronic applications including lasing, photodiodes, light emitting diodes, and solar cells. Understanding the role that exciton diffusion plays in organic solar cells is crucial to understanding the recent rise in power conversion efficiencies brought about by non-fullerene acceptor molecules (NFAs). Established methods for quantifying exciton diffusion lengths in organic semiconductors require specialized equipment designed for measuring high-resolution time-resolved photoluminescence (TRPL). In this article we introduce an approach, named pulsed-PLQY, to determine the diffusion length of excitons in organic semiconductors without any temporal measurements. Using a Monte-Carlo model the dynamics within a thin film semiconductor are simulated and the results are analysed using both pulsed-PLQY and TRPL methods. It is found that pulsed-PLQY has a larger operational window and depends less on the excitation fluence than the TRPL approach. The simulated results are validated experimentally on a well understood organic semiconductor, after which pulsed-PLQY is used to evaluate the diffusion length in a variety of technologically relevant materials. It is found that the diffusion lengths in NFAs are much larger than in the benchmark fullerene and that this increase is driven by an increase in diffusivity. This result helps explain the high charge generation yield in low-offset state-of-the-art NFA solar cells.

\end{abstract}
\maketitle

\clearpage
\section{Introduction}
Organic semiconductors have shown great promise as potential materials for light emitting diodes\cite{kalyani2012}, photodetectors\cite{de2017}, lasers\cite{chenais2012}, and solar cells\cite{armin2021}. Organic solar cells (OSCs), based on organic semiconductors, are a promising solar harvesting technology for many applications such as indoor and stand-alone power, due to a number of advantageous features such as low embodied energy processing, earth abundant constituent materials, suitability for flexible form factor, and tunable optoelectronic properties\cite{cutting2016,davy2017,armin2021}. Traditional OSCs are fabricated by blending a polymer donor and a fullerene based acceptor semiconductor in solution and depositing to form a bulk heterojunction solar cell (BHJ). Due to the low dielectric constants of organic semiconductors ($\epsilon_r\sim$ 2-4), absorption of photons does not directly result in free charge carriers. Rather, bound electron-hole pairs localized to either the donor or acceptor phase known as excitons are the primary photo-excited species\cite{fu2021}. To create free charge carriers, excitons generated in the bulk of either phase must first diffuse to the donor:acceptor interface before they decay. At the interface, an exciton generated in the donor (acceptor) phase can form a charge-transfer (CT) state by transferring the electron (hole) to the acceptor (donor) phase. This is thought to be the primary route for the generation of free charges and is therefore central to the photovoltaic process in OSCs. The transfer of an electron (hole) from the donor (acceptor) to the acceptor (donor) molecule in these cells is typically considered to be driven by an energetic offset between the lowest unoccupied (highest occupied) molecular orbital, LUMO (HOMO), levels of the two materials, known as type I (II) charge generation\cite{armin2014,stoltzfus2016,deibel2010,armin2021}.

The recent advent of non-fullerene acceptor (NFA) based OSCs have propelled power conversion efficiencies consistently above 15\% and as high as 18.2\%\cite{yuan2019,cui2020,qin2020,liu2020,liu22020}. Low-offset state-of-the-art NFA BHJs, such as the benchmark PM6:Y6 system (for a list of the chemical definitions see supplemental material \cite{SUPP}), have minimal offset between the HOMO levels of the donor and acceptor molecule\cite{yuan2019,chen2017}. Despite this apparent disadvantage, low-offset NFA systems show dramatic increases in power conversion efficiency compared to their fullerene predecessors, brought about by increases in short-circuit current and high charge generation yield\cite{xiao2017,yuan2019,cui2020,qin2020,wang2021}. It has been conjectured that the lack of driving force is compensated for by an increase in exciton lifetime or diffusivity, amalgamated into diffusion length, increasing the attempt frequency of CT-state formation\cite{sajjad2020,classen2020}. Other evidence implies F\"{o}rster energy transfer allows for either type I or II charge generation to occur for bulk excitons generated in either phase\cite{karuthedath2021,forster1948}. Thorough understanding of exciton dynamics within the donor and acceptor lattices and exciton kinetics at the donor:acceptor interface is essential to explain the efficient exciton dissociation in low-offset NFA solar cells. Contributing to this understanding is the motivation behind the work presented here.

\begin{figure*}
\includegraphics[scale=0.9]{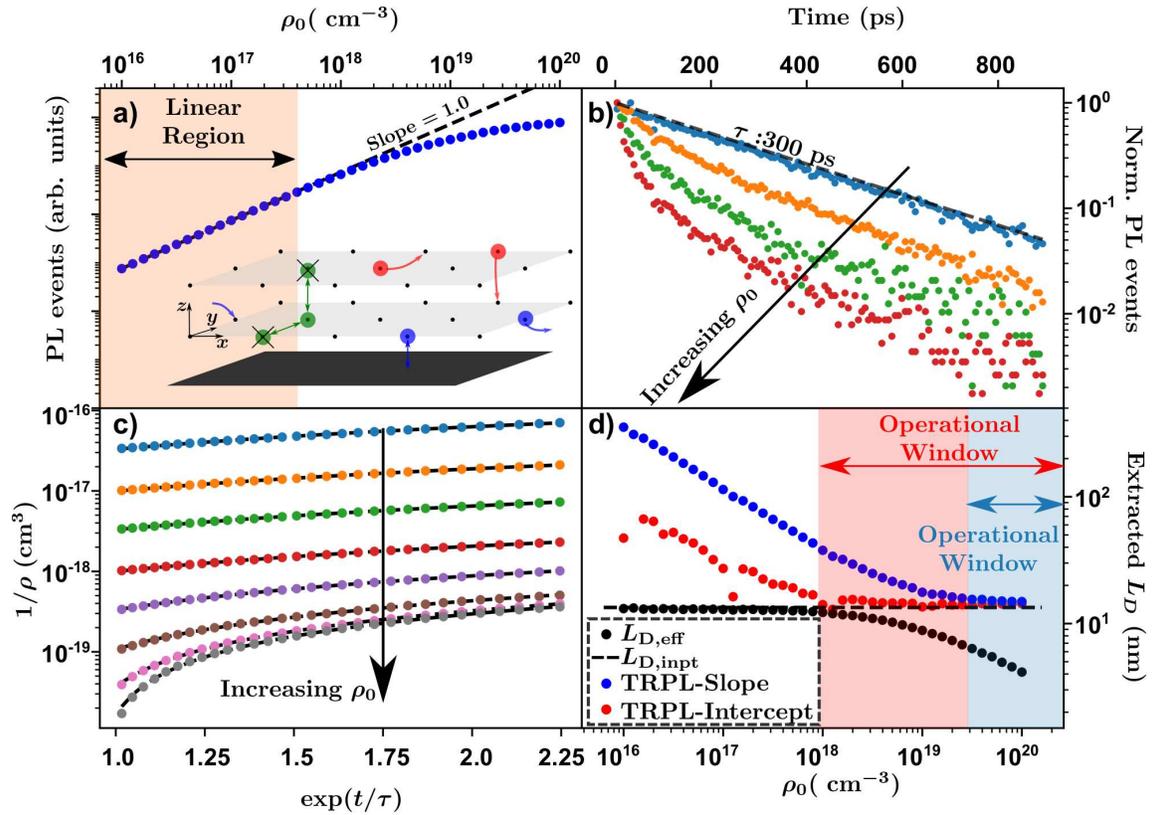}
\caption[]{Simulated (a) PL events as a function of initial excitation density, (inset) cartoon depiction of various pathways for excitons to evolve in the simulation. Red circles and arrows indicate hopping transport along major axes, blue circles and arrows indicate reflecting and periodic boundary conditions, green circles and arrows indicate exciton-exciton annihilation. Simulated (b) TRPL-decays and as a function of time and (c) linearized TRPL decays for selected initial excitation densities (circles) and corresponding fit lines (black dashed lines), note that the axis in panel (c) are log-lin. (d) Diffusion lengths extracted through TRPL-linearization (blue and red circles), calculated from Eq.~\ref{equation:simulatedLD} (black circles), or calculated from simulation parameters (black dashed line).}
\label{figure:simulationTRPL}
\end{figure*}

While exciton lifetimes can be discerned from time-resolved photoluminescence (TRPL) measurements, measuring the exciton diffusivity, identified by the diffusion coefficient, is not as straightforward. Exciton diffusion coefficients in organic semiconductors can be determined through a variety of optical methods including fluorescence volume\cite{ward2012,mikhnenko2012,wang2011} or bi-layer\cite{
theander2000,luhman2011,mikhnenko2008,haugeneder1999,mikhnenko2009,scully2006,markov2005,shaw2008,ward2012} quenching, and exciton-exciton annihilation (EEA)\cite{lewis2006,engel2006,shaw2008,cook2009,cook2010,wang2011,long2017,zhang2019,sajjad2020,park2021} measurements. A robust quenching experiment requires the fabrication of multiple films with varying thickness, detailed measurement of optical constants, a precise understanding of both the heterostructure and the quenching mechanism, and in the case of steady-state measurements the absolute value of the photoluminesence quantum yield ($\eta_\text{PL}$). In contrast, EEA approaches require only one film with spectrally understood absorbance, do not require absolute $\eta_\text{PL}$ measurements, and are less sensitive to both optical interference effects and long-range quenching mechanisms such as F\"{o}rster energy transfer\cite{mikhnenko2015}. Futher, EEA studies provide additional insight into exciton dynamics by observing exciton-exciton interactions at high excitation densities. This ancillary information allows for the determination of the annihilation coefficient, important to the field of organic lasing, and the so called exciton capture radius, defined as the average distance at which excitons annihilate.

In the majority of experimental reports employing EEA the annihilation coefficient is measured through linearizing TRPL data at high excitation densities, while the capture radius is assumed to be on the order of the molecular spacing, allowing for the calculation of the low-density diffusion length\cite{shaw2008,lewis2006,cook2009,cook2010}. In some studies authors further compare diffusion length and annihilation coefficient from two independent methods to gain insight into the capture radius, however; this requires the assumption of a perfect quenching molecule or interface\cite{sajjad2020,shaw2008,long2017}. Still others use global fitting techniques, including the capture radius as either a fitting parameter or an assumed value\cite{wang2011,sajjad2020,park2021}. Recent studies found the $d_{100}$ lamellar spacing measured by grazing-incident wide-angle X-ray scattering experiments (GIWAX) to be a close estimation of the capture radius\cite{sajjad2020}. Although a reliable method for determining the capture radius is yet to be agreed upon, EEA has been a useful tool in determining the exciton diffusion length of many organic semiconductors\cite{mikhnenko2015}.

In this work, we explore the limitations of EEA experiments by evaluating the density dependence of the diffusion length extracted from TRPL-linearization using a Monte-Carlo hopping model. Then an alternative EEA approach which does not require any temporal measurements, termed pulsed-PLQY, is proposed and demonstrated via the same Monte-Carlo simulations. The use of Monte-Carlo simulations allows for the identification of operational windows, defined as the range of densities over which each analysis of the simulated kinetics reproduces the input diffusion length. It is found that, even in the ideal case, pulsed-PLQY has a larger operational window and is less sensitive to the choice of initial exciton density compared to the traditional TRPL-linearization technique. TRPL-linearization and pulsed-PLQY are performed on the well studied model system P3HT and compared. Both experiments reproduce the trends predicted by the simulations. Overall it is found that pulsed-PLQY is less dependent on the excitation fluence, is faster, easier, and requires less specialized equipment than the traditional EEA measurement techniques. Finally, pulsed-PLQY is used to measure the annihilation coefficient, diffusion length, and diffusion coefficient in various organic semiconductors. It is found that diffusion lengths in NFA organic semiconductors are longer than those found in fullerene acceptors and that this difference is driven by an increase in diffusivity.

\section{Theoretical Background}
Singlet-singlet exciton annihilation can occur when two excitons interact with each other, typically assumed to be on neighbouring molecules. The result of this interaction is an exciton with excess energy, which quickly relaxes to the lowest excited-state, and one non-radiative decay event to the ground state. The rate equation for the density of excitons ($\rho)$ in an organic semiconductor is determined by the sum of this second order non-radiative decay and the first order natural decay
\begin{align}
\frac{d\rho(t)}{dt} & = -\frac{\rho(t)}{\tau} - \gamma \rho^2(t)\label{equation:rate}
\end{align}
where $t$ is the time, $\tau$ is the natural (low-density) lifetime for singlet excitons, and $\gamma$ is the exciton-exciton annihilation coefficient. When the process of annihilation is diffusion-limited $\gamma$ can be related to the diffusion constant ($D$) in the film as\cite{chandrasekhar1943}
\begin{align}
\gamma &= 4\pi DR_0\label{equation:gamma}
\end{align}
where $R_0$ is the capture radius. Although, interactions that result in quenching of both excitons and lead to a value for $\gamma$ twice what is derived here have been proposed\cite{engel2006}, previous experimental reports by other authors have confirmed that the primary quenching mechanism in organic semicondcutors is accurately described by Eqs.~\ref{equation:rate} and \ref{equation:gamma} \cite{lewis2006,shaw2008,cook2009,cook2010,mikhnenko2015,long2017,zhang2019,sajjad2020}.

The solution to Eq.~\ref{equation:rate} is
\begin{align}
\rho(t) = \frac{\rho_0\exp\left(-t/\tau\right)}{1+\gamma\rho_0\tau\left[1-\exp\left(-t/\tau\right)\right]}\label{equation:solution}
\end{align}
where $\rho_0$ is the initial exciton density (at $t=0$). Eq.~\ref{equation:solution} can be linearized as
\begin{align}
\frac{1}{\rho(t)} &= \left[\frac{1}{\rho_0} + \gamma\tau\right]\exp\left(t/\tau\right) - \gamma\tau\label{equation:linear}
\end{align}
allowing for $\gamma$ to be obtained from either slope or intercept of a $1/\rho$ vs $\exp(t/\tau)$ plot, assuming $\tau$ is known. Finally, the annihilation coefficient can be related to the low-density diffusion length ($L_D$), through Eq.~\ref{equation:gamma}, noting that
\begin{align}
L_D = \sqrt{2nD\tau}\label{equation:LD}
\end{align}
where $n$ is the dimensionality of the diffusion, although some researchers choose to drop the factor of 2 for convenience.

\section{Results and Discussion}
To investigate the limits of TRPL-linearization and to introduce pulsed-PLQY the two methods were simulated using a Monte-Carlo hopping model. The simulations were limited to capture only the relevant physics and to allow for each experimental method to be evaluated under ideal conditions where system parameters, such as the diffusion length, are known and can be compared to the extracted values. To this end, the exciton dynamics within a lattice were simulated over a range of initial excitation densities with a 3D Monte-Carlo model limited to nearest neighbour hopping pairs but including both natural (linear) and exciton-exciton annihilation decay mechanisms (as shown in the inset of Figure \ref{figure:simulationTRPL} (a)). Here, exciton-exciton annihilation is assumed to occur when excitons are within one lattice spacing of one another (corresponding to an $R_0$ equal to the lattice spacing). Although it is possible that exciton-exciton annihilation occurs both over greater distances and as a statistical process, the capture radius is typically understood to be the average distance over which annihilation occurs. Limiting the annihilation to one lattice spacing allows for an absolute calculation of the diffusion length through Eqs.~\ref{equation:gamma} and \ref{equation:LD} (with $R_0$ equal to the lattice spacing) and a direct comparison of the diffusion length calculated from the simulated TRPL-linearization and pulsed-PLQY techniques. Further, the finding that the GIWAX $d_{100}$ spacing is similar to the capture radius experimentally corroborates the view that annihilation, on average, happens between nearest neighbour pairs\cite{sajjad2020}. The lattice spacing ($dx$), temporal step size, and natural lifetime used were 0.775 nm, 1 ps, and 300 ps respectively, corresponding to a diffusion coefficient of $10^{-3}$ $\text{cm}^2/\text{s}$ and an input 3D diffusion length ($L_{D\text{,inpt}}$) of 13.4 nm. The details of the Monte-Carlo Model are outlined in Appendix \ref{appendix:MC}.
 
Figure \ref{figure:simulationTRPL} (a) shows the photoluminescence, determined by the number of natural decay events,  as a function of initial exciton density. A transition between first and second-order response, as expected from Eq.~\ref{equation:rate}, can clearly be seen starting around $3\times 10^{17}$ $\text{cm}^{-3}$ and fully occurring by $3\times 10^{18}$ $\text{cm}^{-3}$. This is reflected in the increased quenching occurring on sub-100 ps time scale in the selected TRPL curves shown in Figure \ref{figure:simulationTRPL} (b), typical of exciton-exciton annihilation experiments\cite{lewis2006,engel2006,shaw2008,cook2009,cook2010,wang2011,long2017,zhang2019,sajjad2020,park2021}.

 To explore the limits of Eq.~\ref{equation:linear} the simulated TRPL data was analyzed, through TRPL-linearization, to obtain the diffusion length and compare with $L_{D\text{,inpt}}$. In this technique the lowest density TRPL data is fit to obtain $\tau$ as shown by the black dashed line in Figure \ref{figure:simulationTRPL} (b). Using this value of $\tau$ each TRPL decay is linearized according to Eq.~\ref{equation:linear} and fit to a line, as shown for the selected curves in Figure \ref{figure:simulationTRPL} (c). The slope and intercept are used to obtain values of $\gamma$ for each initial density and the diffusion length is calculated from Eqs.~\ref{equation:gamma} and \ref{equation:LD}, using $R_0 = dx$. This is equivalent to the procedure for fitting experimental data.

The black dashed line in Figure \ref{figure:simulationTRPL} (d) shows the input diffusion length ($L_{D\text{,inpt}}$) and the black circles indicate the effective diffusion length ($L_{D\text{,eff}}$, calculated from simulated exciton dynamics, see appendix \ref{appendix:MC}) of the simulated excitons as a function of initial density. At low-densities $L_{D\text{,eff}}$ predicts $L_{D\text{,inpt}}$, however, $L_{D\text{,eff}}$ increasingly underestimates $L_{D\text{,inpt}}$ with increasing $\rho_0$. This is due to an increase in exciton-exciton annihilation occurring at large $\rho_0$, where, on average, excitons have a shorter effective lifetime (and therefore move, on average, less distance) than those at low-densities. It is therefore expected that $L_{D\text{,eff}}$ will decrease with increasing $\rho_0$.

\begin{figure}
\includegraphics[scale=0.8]{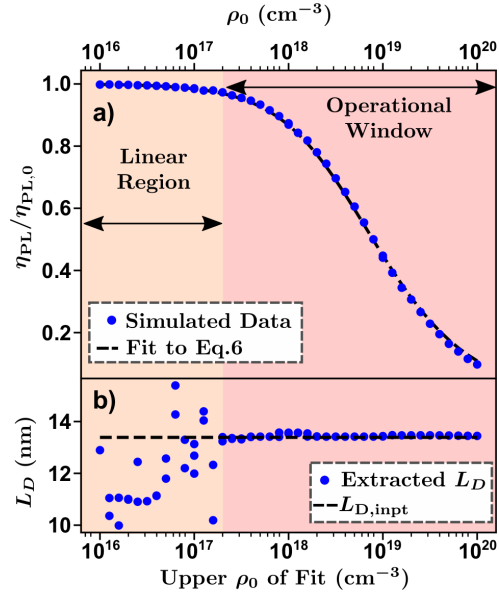}
\caption[]{(a) Simulated $\eta_\text{PL}$ as a function of initial excitation density (blue circles) and fit to Eq.~\ref{equation:PLQY} (black dashed line). (b) Exciton diffusion length extracted from data in panel (a) as a function of the upper limit to the fit (blue circles) and input diffusion length (black dashed line).}
\label{figure:simulationPLQY}
\end{figure}

In contrast to this, the blue and red circles in Figure \ref{figure:simulationTRPL} (d), respectively, show the diffusion lengths extracted from the slope and intercept of the fit to the linearized simulated TRPL data. The extracted diffusion lengths reproduce $L_{D\text{,inpt}}$ at high $\rho_0$, where the second-order non-radiative decay dominates Eq.~\ref{equation:rate}. However, the diffusion length (or $\gamma$) extracted from the slope increasingly overestimates the input diffusion length with decreasing $\rho_0$. Similar reduction in extracted annihilation coefficient (as measured by high density TRPL experiments) with increasing excitation density has been observed by other researchers in previous studies on organic semiconducting polymers\cite{lewis2006}.

 The apparent dependence of the slope on $\rho_0$ limits the range over which the true value of $\gamma$ (and hence $L_D$) can be extracted and the extracted value converges to the expected low-density diffusion length only when $\gamma\tau \gg 1/\rho_0$. The operational window is defined as the range of densities over which this technique reliably reproduces the low-density diffusion length. The operational window for extraction from the slope is indicated by the blue shading in Figure \ref{figure:simulationTRPL} (d). On the other hand, according to Eq.~\ref{equation:linear}, the intercept does not depend on $\rho_0$. This is reflected in the larger operational window for the diffusion length extracted from the intercept (red shaded region in Figure \ref{figure:simulationTRPL} (d)). However, in the limit where $\rho_0\gamma\tau\ll 1$ Eq.~\ref{equation:solution} is reduced to a single exponential decay and one cannot expect to extract any second-order information, reflected in the increasing overestimation of the diffusion length as the density approaches the linear regime.
 
Consequently, accurate quantification of the diffusion length (or $\gamma$) from TRPL-linearzation requires careful analysis of the range of excitation densities used. Implementation of this experimental procedure requires a femtosecond pulsed laser source to inject (and subsequently let evolve) the initial exciton density, accurate measurement of the natural lifetime, and specialized equipment capable of measuring the quenching kinetics seen in Figure \ref{figure:simulationTRPL} (b), which occur on the order of picoseconds in organic semiconductors\cite{engel2006,shaw2008,cook2009,cook2010,wang2011,zhang2019,sajjad2020}. Furthermore, as described above, the operational window for determining the diffusion length (or $\gamma$) of these materials occurs at high excitation density where the quenching is fastest requiring increasing resolution to resolve. Besides which, not all organic semiconductors are stable at high excitation densities and may undergo photo- or thermal-oxidation at sufficiently high excitation fluences.

\begin{figure*}
\includegraphics[scale=0.96]{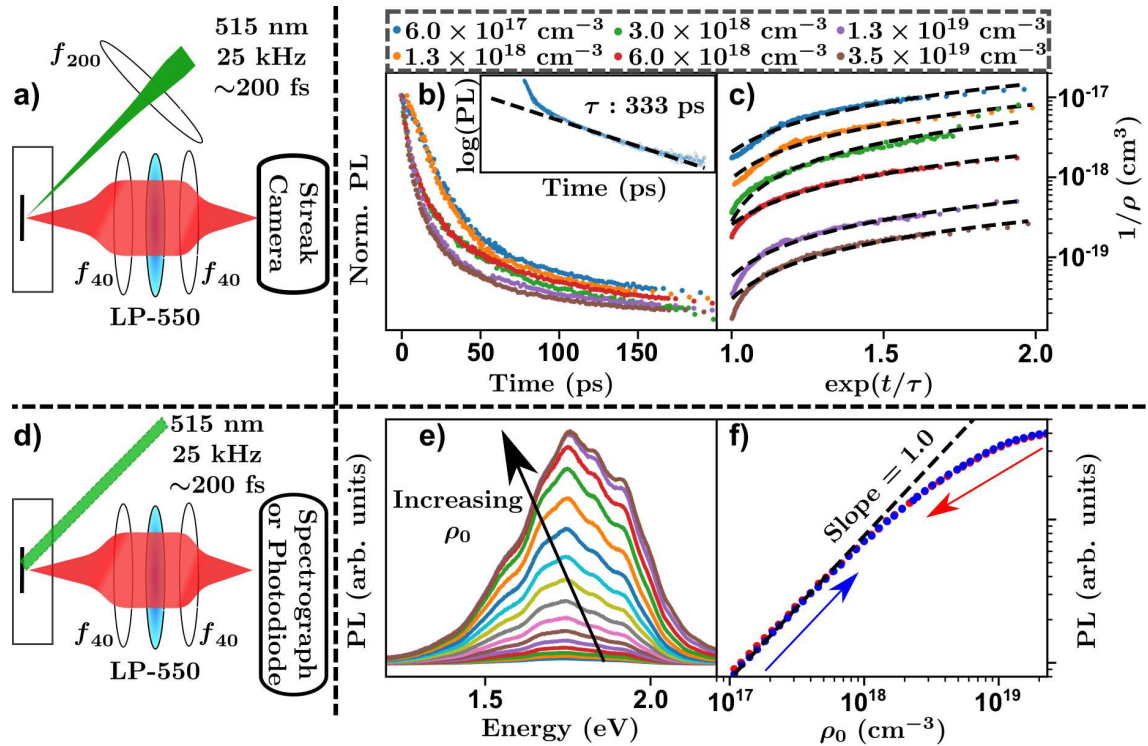}
\caption[]{(a) TRPL measurement apparatus. (b) Normalized TRPL decays as a function of time and (c) linearized TRPL decays for various initial excitation densities. (d) Pulsed-PLQY measurement apparatus. (e) P3HT spectra for selected excitation densities. (f) P3HT peak PL counts as a function of initial excitation density. $f_{200/40}$ indicate focal lengths of lenses, LP-long pass filter.}
\label{figure:expApparatus}
\end{figure*}

 To circumvent the apparent dependence of extracted values on the choice of initial density, eliminate the requirements for measuring the natural lifetime, lower the equipment specialization, and increase the speed of the measurement one can measure and analyze these kinetics in an alternative way by considering the normalized photoluminescence quantum efficiency ($\eta_\text{PL}$) as a function of initial density. The total number of excitons that recombine via natural decay in accordance with Eq.~\ref{equation:rate} is $\rho_\tau = \int_0^\infty\rho(t)/\tau dt$. On the other hand, the photoluminescence quantum efficiency is given by $\eta_\text{PL} = \rho_\tau\eta_\text{PL,0}/\rho_0$ where $\eta_\text{PL,0}$ is the low density photoluminescence quantum efficiency. From Eq.~\ref{equation:solution}, the photoluminescence quantum efficiency is then obtained as

\begin{align}
\eta_\text{PL}\left(\rho_0\right)  &= \eta_\text{PL,0}\frac{\ln\left[1+\rho_0\gamma\tau\right]}{\rho_0\gamma\tau}\label{equation:PLQY}
\end{align}
Therefore, by measuring $\eta_\text{PL}/\eta_\text{PL,0}$ as a function of $\rho_0$ one can fit to obtain the product of $\gamma\tau$ and either: (i) calculate the the diffusion length directly using Eqs.~\ref{equation:gamma} and \ref{equation:LD} without the need for any temporal measurements, or (ii) measure the natural lifetime to calculate $\gamma$ (or $D$).

To evaluate the validity of this procedure, $\eta_\text{PL}$ was calculated from the results of the Monte-Carlo simulations as the ratio of natural decay events to the initial number of excitons in the lattice. Figure \ref{figure:simulationPLQY} (a) shows the results of this analysis, the blue circles indicate $\eta_\text{PL}$ calculated for each initial density. The exciton-exciton annihilation is evident in the decrease in $\eta_\text{PL}$ with increasing $\rho_0$, as observed experimentally by other researchers in studies on metal-insulator-semiconductor heterostructures\cite{kumar2014,linardy2020}. The linear region marked in Figure \ref{figure:simulationTRPL} (a) is reproduced in Figure \ref{figure:simulationPLQY} as a guide.

The simulated normalized $\eta_\text{PL}$ is fit to Eq.~\ref{equation:PLQY}, as shown by the black dashed line and the diffusion length is calculated from the extracted product of $\gamma\tau$ through Eqs.~\ref{equation:gamma} and \ref{equation:LD}. Figure \ref{figure:simulationPLQY} (b) shows the calculated diffusion length as a function of the upper limit to the fitting. The operational window is defined as the densities that, when used as the upper limit to the fitting, reproduce the low-density diffusion length. The operational window for pulsed-PLQY is shown by the red shaded region in Figure \ref{figure:simulationPLQY} (a) and (b). The operational window for pulsed-PLQY is larger and begins at lower densities that those of TRPL-linearization and, most importantly, the confidence in the extracted value increases with increasing number of initial densities used. Whereas, in TRPL-linearization increases in the number of measurements made at varying initial densities may decrease the confidence of the extracted values, depending on the choices of $\rho_0$.

Actual realization of this experiment requires the same femtosecond pulsed laser as with TRPL-linearization, however, the light-collection can be accomplished with non-specialized equipment operating at a quasi-steady state. Further, the exciton diffusion length can be calculated directly from the product $\gamma\tau$ without the need for any temporal measurements, and $\gamma$ (or $D$) can be determined without the need for a high resolution TRPL apparatus.

To validate these simulations TRPL-linearlization and pulsed-PLQY were carried out on a P3HT thin film and the results compared. Figure \ref{figure:expApparatus} (a) shows the experimental TRPL-linearization apparatus (detailed in Appendix \ref{appendix:Exp}). Figure \ref{figure:expApparatus} (b) shows the resulting normalized TRPL decays, where the expected time-independent quenching dynamics can be seen, akin to other studies on similar materials\cite{shaw2008,lewis2006,cook2009,cook2010,sajjad2020,long2017,wang2011,park2021}. The natural lifetime was measured by fitting the decay of the lowest initial density to the linear region (where exciton-exciton annihilation is minimal) on a log-lin scale as shown in the inset of Figure \ref{figure:expApparatus} (b). The natural lifetime was used to linearize all the TRPL decays in Figure \ref{figure:expApparatus} (b) according to Eq.~\ref{equation:linear}, as shown in Figure \ref{figure:expApparatus} (c). The black dashed lines indicate the fits used to extract the annihilation coefficient from the intercept, as this was shown to be the more accurate measure. 

\begin{figure}
\includegraphics[scale=0.8]{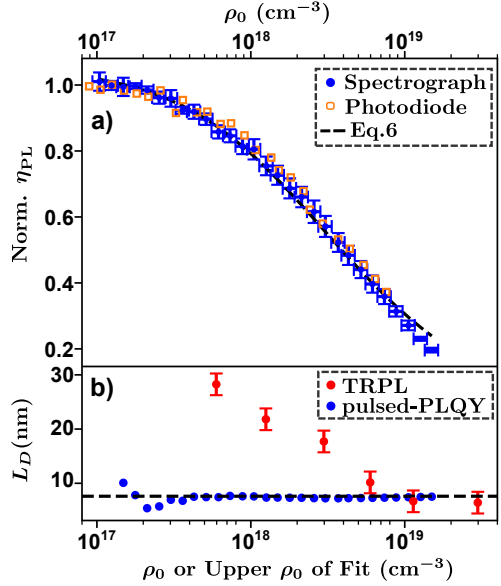}
\caption[]{P3HT (a) pulsed-PLQY data collected using spectrograph (blue circles) and lock-in (orange squared) detection, and fitting of spectrograph data to Eq.~\ref{equation:PLQY} (black line). (b) Diffusion lengths extracted from the intercept of TRPL-linearlization as a function of initial density (red circles) and pulsed-PLQY as a function of upper limit to the fit.}
\label{figure:expPLQY}
\end{figure}

Figure \ref{figure:expApparatus} (d) shows the apparatus used for the pulsed-PLQY technique (detailed in Appendix \ref{appendix:Exp}). Selected spectra obtained from the same P3HT thin film are shown in Figure \ref{figure:expApparatus} (e). The peak values of the spectra are plotted against the excitation density in Figure \ref{figure:expApparatus} (f). The transition from the linear to sub-linear response can be clearly seen occurring around the same values as those in Figure \ref{figure:simulationTRPL} (a). To ensure that there was no degradation of the sample during these measurements both forward (blue circles-from low to high density) and reverse (red circles-from high to low-density) scans were performed consecutively. To calculate the relative $\eta_\text{PL}$ the peak value of the photoluminescence data was divided by the associated $\rho_0$ and normalized to the $\eta_\text{PL}$ of the lowest density points, as shown by the blue circles in Figure~\ref{figure:expPLQY} (a). To exemplify the simplicity of this measurement a second set of data was collected where the imaging spectrograph was replaced with a silicon photodiode and the voltage response was used to calculate the relative $\eta_\text{PL}$. This second data set is shown as the orange squares in Figure \ref{figure:expPLQY} (a). Although the use of a silicon photodiode does not provide spectral information, the relative $\eta_\text{PL}$, and therefore $\gamma\tau$ (or $L_D$), can be measured with a further reduction in equipment specialization. The $\eta_\text{PL}$ data collected with the spectrograph are fit to Eq.~\ref{equation:PLQY}, shown by the black dashed line, and the extracted product of $\gamma\tau$ is used, with the $d_{100}$ spacing as an estimate of $R_0$, to calculate the diffusion length via Eqs.~\ref{equation:gamma} and \ref{equation:LD}. 

The results of the experiments on P3HT are summarized in Figure \ref{figure:expPLQY} (b) which shows the 1D diffusion length extracted from TRPL-linearization as a function of initial density and pulsed-PLQY as a function of upper limit to the fit. As predicted by the simulations the diffusion length extracted from TRPL-linearization (red circles) decreases with increasing initial density. The diffusion length measured by pulsed-PLQY (blue circles) converges to a constant value with increasing upper fitting limit. These results verify the simulations and demonstrate that pulsed-PLQY is a viable and accurate alternative measurement technique for measuring diffusion lengths without the need for any temporal measurements, or the annihilation coefficient without the need for high resolution TRPL equipment. 

The product of $\gamma\tau$ found from pulsed-PLQY was $(7.2\pm 0.2)\times10^{-19}$ $\text{cm}^{3}$ (for a detailed description of error analysis see supplemental material\cite{SUPP}), leading to a 1D diffusion length of $8\pm1$ nm, comparable to literature values\cite{shaw2008,mikhnenko2015}. Using the natural lifetime, found to be $330 \pm 10$ ps, a value of $\gamma=(2.15\pm 0.08)\times 10^{-9}$ $\text{cm}^{-3}$ was found for the annihilation coefficient, slightly lower than previous values extracted from TRPL-linearization\cite{shaw2008}. However, as exemplified by the simulations, a slightly lower value of $\gamma$ is expected from pulsed-PLQY than from TRPL-linearization. Finally, the diffusion coefficient was calculated from $\gamma$, and $R_0$ according to Eq.~\ref{equation:gamma} and found to be $(1.0\pm 0.3)\times 10^{-3}$ $\text{cm}^2/$s, in agreement with previous values from quenching experiments\cite{shaw2008,wang2011}.

\begin{table*}
\caption[]{Material parameters extracted from the pulsed-PLQY fittings ($\gamma\tau$ and $L_D$), low-density TRPL fittings ($\tau$), and subsequently calculated ($\gamma$, and $D$) for various organic semiconductors. See supplemental material for all measurement data\cite{SUPP}. $R_0$ is assumed to be the $d_{100}$ spacing taken from the provided references, the error in $R_0$ is assumed half the value given by the $d_{100}$ spacing. $\dagger \text{PC}_{60}\text{BM}$ estimate for $R_0$ is taken as 1 nm as GIWAX $d_{100}$ does not give intermolecular distance in this case.}
\begin{tabular}{l|c|c|c|c|c|c|c}
\hline
\textbf{Material}&\textbf{\begin{tabular}[c]{@{}c@{}}$\boldsymbol{\gamma\tau}(\boldsymbol{\times10^{-18}}$)\\($\text{cm}^\textbf{3}$)\end{tabular}}&\textbf{\begin{tabular}[c]{@{}c@{}}$\boldsymbol{R_0}$\\(nm)\end{tabular}}&\textbf{\begin{tabular}[c]{@{}c@{}}$\boldsymbol{L_D^{1D}}$\\(nm)\end{tabular}}&\textbf{\begin{tabular}[c]{@{}c@{}}$\boldsymbol{L_D^{3D}}$\\(nm)\end{tabular}}&\textbf{\begin{tabular}[c]{@{}c@{}}$\boldsymbol{\tau}$\\(ps)\end{tabular}}&\textbf{\begin{tabular}[c]{@{}c@{}}$\boldsymbol{\gamma}$($\boldsymbol{\times10^{-9}}$)\\($\text{cm}^\textbf{3}/\text{s}$)\end{tabular}}&\textbf{\begin{tabular}[c]{@{}c@{}}\textbf{D}($\boldsymbol{\times10^{-4}}$)\\($\text{cm}^\textbf{2}/\text{s}$)\end{tabular}}\\\hline
PM6&0.088$\pm$0.002&2.1\cite{yuan2019}&2.6$\pm$0.3&4.5$\pm$0.3&64&1.4$\pm$0.2&5$\pm$1\\
P3HT-Regiorandom&0.134$\pm$0.02&1.0\cite{deshmukh2018}&5$\pm$1&8$\pm$1&175&0.76$\pm$0.05&6$\pm$3\\
PTB7-Th&0.144$\pm$0.004&2.4\cite{deshmukh2018}&3.1$\pm$0.3&5.4$\pm$0.3&123&1.17$\pm$0.09&3.9$\pm$0.9\\
PCDTBT&0.329$\pm$0.005&1.6\cite{wong2017}&5.7$\pm$0.9&9.9$\pm$0.9&203&1.62$\pm$0.08&8$\pm$2\\
$\text{PC}_{60}\text{BM}$&0.52$\pm$0.01&$1.0^\dagger$&9$\pm$2&23$\pm$3&777&0.65$\pm$0.02&5$\pm$2\\
P3HT-Regioregular&0.72$\pm$0.02&1.0\cite{chu2016}&8$\pm$1&14$\pm$1&333&2.15$\pm$0.08&10$\pm$3\\
ITIC (NFA)&1.23$\pm$0.03&1.6\cite{mai2018}&11$\pm$2&19$\pm$2&120&10.3$\pm$0.9&50$\pm$10\\
IT4F (NFA)&2.54$\pm$0.07&1.8\cite{ma2020}&15$\pm$2&26$\pm$2&205&12.4$\pm$0.7&60$\pm$10\\
Y6 (NFA)&8.5$\pm$0.2&1.5\cite{yuan2019}&30$\pm$5&51$\pm$5&799&10.6$\pm$0.3&60$\pm$10\\
BTP-eC9 (NFA)&10.6$\pm$0.3&1.6\cite{cui2020}&33$\pm$5&56$\pm$5&389&27$\pm$1&140$\pm$40\\
\hline
\end{tabular}
\label{table:results}
\end{table*}

After pulsed-PLQY was established it was used to evaluate $\gamma\tau$ and the subsequent 1D and 3D diffusion lengths in a collection of organic semiconductors, including several technologically relevant NFAs used in state-of-the-art organic solar cells. Further, the natural lifetime was measured, at low excitation density, to evaluate the annihilation and diffusion coefficients. These results are summarized in Table \ref{table:results}. The NFAs (such as ITIC, IT4F, Y6, and BTP-eC9) were found to have much longer diffusion lengths than the benchmark fullerene acceptor $\text{PC}_{60}\text{BM}$. Measurement of the natural lifetime shows that the increase in diffusion length is driven by increased diffusivity. Increases in diffusivity would allow bulk excitons to form interfacial excitons many times over, increasing the probability of CT-state formation regardless of HOMO offset. However, this evidence does not rule out F\"{o}rster energy transfer playing a significant role in charge generation. The F\"{o}rster radii, defined as the range at which natural decay and long range transfer rates are equivalent, is $<5$ nm in BHJs made from these materials\cite{karuthedath2021}. If F\"{o}rster energy transfer were to occur, suppressing one charge generation channel, increased diffusion lengths would still increase charge generation by increasing the number of bulk excitons that diffuse close enough to the interface for long-range F\"{o}rster energy transfer to efficiently transfer the exciton to the other phase. Alternatively, for excitons that have undergone F\"{o}rster energy transfer to the bulk of the other phase increased diffusivity would increase the chance of forming a interfacial exciton from the new phase.

\section{Conclusions}
In conclusion, pulsed-PLQY was introduced as an alternative approach to established exciton-exciton annihilation techniques. A Monte-Carlo hopping model was used to validate the experimental technique and the subsequent results on P3HT agree with the models predictions. Pulsed-PLQY was then used to determine the diffusion lengths of various organic semiconductors and the results show a dramatic increase in diffusion length in NFA materials, primarily driven by increased diffusivity. These results support the assertation that increases in diffusion constant contribute to the charge generation efficiency in low-offset NFA based organic solar cells. Pulsed-PLQY will contribute to the field of organic photovoltaics specifically, and organic optoelectronics more broadly, by providing an alternative technique to measure the diffusion length in organic semiconductors without the need for temporal measurements, or to measure annihilation coefficients without the need for a high resolution TRPL apparatus.

\begin{acknowledgments}
This work was supported by the Welsh Government's S\^er Cymru II Program through the European Regional Development Fund, Welsh European Funding Office, and Swansea University strategic initiative in Sustainable Advanced Materials. A.A. is a S\^er Cymru II Rising Star Fellow and P.M. is a S\^er Cymru II National Research Chair. This work was also funded by UKRI through the EPSRC Program Grant EP/T028511/1 Application Targeted Integrated Photovoltaics. D.B.R acknowledges the support of the Natural Sciences and Engineering Research Council of Canada (NSERC), [PGSD3-545694-2020]. The authors acknowledge the support of the Supercomputing Wales project, which is part-funded by the European Regional Development Fund (ERDF) via the Welsh Government. The authors thank Dr. Nasim Zarrabi for her insight and fruitful discussions.
\end{acknowledgments}

\appendix
\section{Monte-Carlo Model}\label{appendix:MC}
With the aim of modelling the dynamics within an organic semiconductor a 3D Monte-Carlo hopping model was invoked\cite{kaiser2018}. First a lattice of sufficient volume ($V$) to accommodate the input density was created with spacing between each point $dx$. The lattice is randomly populated with $N$ excitons (where $N = \rho_0V$). Once the lattice has been initialized the simulation is evolved in time with a temporal step-size $dt$, defining the diffusion coefficient as $D = dx^2/6dt$.

The inset in Figure \ref{figure:simulationTRPL} (a) (main text) indicates the various pathways an exciton in the system can evolve. The green circles and arrows indicate annihilation, where any exciton within one lattice spacing of another will decay non-radiatively via exciton-exciton annihilation, such that $R_0$ in Eq.~\ref{equation:gamma} equals $dx$. The red circles and arrows in the inset of Figure \ref{figure:simulationTRPL} (a) indicate an exciton hopping to an adjacent lattice site along the major axes. The choice of site to hop to is given by the Gillespie algorithm with equal probability for movement in each direction but limited to nearest neighbour pairs, and a maximum hopping rate given by the inverse of the temporal step size\cite{gillespie}. The amount of time the exciton occupies this new site (the dwell-time, $\tau_\text{Dwell}$) is given by the total hopping rate away from the site ($\Gamma_i$) as $\tau_\text{Dwell} = T/\Gamma_i$, where $T$ is a pseudo-random number generated from an exponential distribution with unit expectation value and unit variance.

To simulate a thin film deposited on glass periodic boundary conditions on the $x$ and $y$ axis and reflecting boundary conditions on the $z$ axis were implemented, as indicated by the blue circles and arrows in the inset of Figure \ref{figure:simulationTRPL} (a). Further, the size of the $z$ dimension was held at 50 nm. The final pathway for an exciton to evolve is natural decay. The probability of natural decay is given by the input natural lifetime ($\tau$) as $dt/\tau$. Therefore, after each dwell time a pseudo-random number on a unit interval is generated, an exciton decays naturally when this is smaller than the decay probability.

Once the system has sufficiently evolved the effective diffusion length can be found from the Euclidean distance travelled by each exciton ($L_i$)
\begin{align}
L_{D\text{,eff}} = \sqrt{\frac{\sum_0^N \left(L_i\right)^2}{N}}\label{equation:simulatedLD}
\end{align}
which will converge to the input diffusion length ($L_{D\text{,input}}=\sqrt{6D\tau}$) in the low-density limit.

\section{Experimental}\label{appendix:Exp}
\noindent\textit{Time Resolved Photoluminescence:} The thin films were held in a cryostat (Linkam LTS420) under a constant flow of nitrogen gas to prevent photo-oxidation\cite{gerischer1991}, and an initial density of excitons ($\rho_0 = P_\text{abs}/\pi \omega^2f_\text{rep}E_\text{ph}d$, $P_\text{abs}$-absorbed laser power, $f_\text{rep}$-reprate and $\omega$-spotsize of the laser, see supplemental material\cite{SUPP,yoshida1976}, $E_\text{Ph}$-incident photon energy, $d$-film thickness) is injected by an ultrafast laser source (Pharos PHM02-2H-3H). The resulting photoluminescence is collimated, filtered to remove scattered pump light, and focused into the streak camera (Hamamatsu C14831).

\noindent\textit{Pulsed-PLQY:} Here the focusing lens is removed from the laser path in order to make a uniform (in-plane) distribution of excitons and the streak camera is replaced with an imaging spectrograph (Hamamatsu C14631-03), or a photo-diode (Thorlabs SM1PD1A) connected to a lock-in amplifier (Stanford Research Systems SR860), with the pump laser chopped at the reference frequency. A complete schematic and description of the pulsed-PLQY apparatus using a photo-diode is included in the supplementary material \cite{SUPP}.

\noindent\textit{Film thickness:} The film thickness was measured from ellipsometry data (J.A. Woollam M-2000 Spectroscopic Ellipsometer).

\bibliography{Pulsed-PLQY}

\begin{thebibliography}{54}%
\makeatletter
\providecommand \@ifxundefined [1]{%
 \@ifx{#1\undefined}
}%
\providecommand \@ifnum [1]{%
 \ifnum #1\expandafter \@firstoftwo
 \else \expandafter \@secondoftwo
 \fi
}%
\providecommand \@ifx [1]{%
 \ifx #1\expandafter \@firstoftwo
 \else \expandafter \@secondoftwo
 \fi
}%
\providecommand \natexlab [1]{#1}%
\providecommand \enquote  [1]{``#1''}%
\providecommand \bibnamefont  [1]{#1}%
\providecommand \bibfnamefont [1]{#1}%
\providecommand \citenamefont [1]{#1}%
\providecommand \href@noop [0]{\@secondoftwo}%
\providecommand \href [0]{\begingroup \@sanitize@url \@href}%
\providecommand \@href[1]{\@@startlink{#1}\@@href}%
\providecommand \@@href[1]{\endgroup#1\@@endlink}%
\providecommand \@sanitize@url [0]{\catcode `\\12\catcode `\$12\catcode
  `\&12\catcode `\#12\catcode `\^12\catcode `\_12\catcode `\%12\relax}%
\providecommand \@@startlink[1]{}%
\providecommand \@@endlink[0]{}%
\providecommand \url  [0]{\begingroup\@sanitize@url \@url }%
\providecommand \@url [1]{\endgroup\@href {#1}{\urlprefix }}%
\providecommand \urlprefix  [0]{URL }%
\providecommand \Eprint [0]{\href }%
\providecommand \doibase [0]{http://dx.doi.org/}%
\providecommand \selectlanguage [0]{\@gobble}%
\providecommand \bibinfo  [0]{\@secondoftwo}%
\providecommand \bibfield  [0]{\@secondoftwo}%
\providecommand \translation [1]{[#1]}%
\providecommand \BibitemOpen [0]{}%
\providecommand \bibitemStop [0]{}%
\providecommand \bibitemNoStop [0]{.\EOS\space}%
\providecommand \EOS [0]{\spacefactor3000\relax}%
\providecommand \BibitemShut  [1]{\csname bibitem#1\endcsname}%
\let\auto@bib@innerbib\@empty
\bibitem [{\citenamefont {Kalyani}\ and\ \citenamefont
  {Dhoble}(2012)}]{kalyani2012}%
  \BibitemOpen
  \bibfield  {author} {\bibinfo {author} {\bibfnamefont {N.~T.}\ \bibnamefont
  {Kalyani}}\ and\ \bibinfo {author} {\bibfnamefont {S.}~\bibnamefont
  {Dhoble}},\ }\bibfield  {title} {\enquote {\bibinfo {title} {Organic light
  emitting diodes: Energy saving lighting technology—a review},}\ }\href@noop
  {} {\bibfield  {journal} {\bibinfo  {journal} {Renew. Sust. Energ. Rev.}\
  }\textbf {\bibinfo {volume} {16}},\ \bibinfo {pages} {2696--2723} (\bibinfo
  {year} {2012})}\BibitemShut {NoStop}%
\bibitem [{\citenamefont {de~Arquer}\ \emph {et~al.}(2017)\citenamefont
  {de~Arquer}, \citenamefont {Armin}, \citenamefont {Meredith},\ and\
  \citenamefont {Sargent}}]{de2017}%
  \BibitemOpen
  \bibfield  {author} {\bibinfo {author} {\bibfnamefont {F.~P.~G.}\
  \bibnamefont {de~Arquer}}, \bibinfo {author} {\bibfnamefont {A.}~\bibnamefont
  {Armin}}, \bibinfo {author} {\bibfnamefont {P.}~\bibnamefont {Meredith}}, \
  and\ \bibinfo {author} {\bibfnamefont {E.~H.}\ \bibnamefont {Sargent}},\
  }\bibfield  {title} {\enquote {\bibinfo {title} {Solution-processed
  semiconductors for next-generation photodetectors},}\ }\href@noop {}
  {\bibfield  {journal} {\bibinfo  {journal} {Nat. Rev. Mater.}\ }\textbf
  {\bibinfo {volume} {2}},\ \bibinfo {pages} {1--17} (\bibinfo {year}
  {2017})}\BibitemShut {NoStop}%
\bibitem [{\citenamefont {Ch{\'e}nais}\ and\ \citenamefont
  {Forget}(2012)}]{chenais2012}%
  \BibitemOpen
  \bibfield  {author} {\bibinfo {author} {\bibfnamefont {S.}~\bibnamefont
  {Ch{\'e}nais}}\ and\ \bibinfo {author} {\bibfnamefont {S.}~\bibnamefont
  {Forget}},\ }\bibfield  {title} {\enquote {\bibinfo {title} {Recent advances
  in solid-state organic lasers},}\ }\href@noop {} {\bibfield  {journal}
  {\bibinfo  {journal} {Polym. Int.}\ }\textbf {\bibinfo {volume} {61}},\
  \bibinfo {pages} {390--406} (\bibinfo {year} {2012})}\BibitemShut {NoStop}%
\bibitem [{\citenamefont {Armin}\ \emph {et~al.}(2021)\citenamefont {Armin},
  \citenamefont {Li}, \citenamefont {Sandberg}, \citenamefont {Xiao},
  \citenamefont {Ding}, \citenamefont {Nelson}, \citenamefont {Neher},
  \citenamefont {Vandewal}, \citenamefont {Shoaee}, \citenamefont {Wang} \emph
  {et~al.}}]{armin2021}%
  \BibitemOpen
  \bibfield  {author} {\bibinfo {author} {\bibfnamefont {A.}~\bibnamefont
  {Armin}}, \bibinfo {author} {\bibfnamefont {W.}~\bibnamefont {Li}}, \bibinfo
  {author} {\bibfnamefont {O.~J.}\ \bibnamefont {Sandberg}}, \bibinfo {author}
  {\bibfnamefont {Z.}~\bibnamefont {Xiao}}, \bibinfo {author} {\bibfnamefont
  {L.}~\bibnamefont {Ding}}, \bibinfo {author} {\bibfnamefont {J.}~\bibnamefont
  {Nelson}}, \bibinfo {author} {\bibfnamefont {D.}~\bibnamefont {Neher}},
  \bibinfo {author} {\bibfnamefont {K.}~\bibnamefont {Vandewal}}, \bibinfo
  {author} {\bibfnamefont {S.}~\bibnamefont {Shoaee}}, \bibinfo {author}
  {\bibfnamefont {T.}~\bibnamefont {Wang}},  \emph {et~al.},\ }\bibfield
  {title} {\enquote {\bibinfo {title} {A history and perspective of
  non-fullerene electron acceptors for organic solar cells},}\ }\href@noop {}
  {\bibfield  {journal} {\bibinfo  {journal} {Adv. Energy. Mater.}\ }\textbf
  {\bibinfo {volume} {11}},\ \bibinfo {pages} {2003570} (\bibinfo {year}
  {2021})}\BibitemShut {NoStop}%
\bibitem [{\citenamefont {Cutting}\ \emph {et~al.}(2016)\citenamefont
  {Cutting}, \citenamefont {Bag},\ and\ \citenamefont
  {Venkataraman}}]{cutting2016}%
  \BibitemOpen
  \bibfield  {author} {\bibinfo {author} {\bibfnamefont {C.~L.}\ \bibnamefont
  {Cutting}}, \bibinfo {author} {\bibfnamefont {M.}~\bibnamefont {Bag}}, \ and\
  \bibinfo {author} {\bibfnamefont {D.}~\bibnamefont {Venkataraman}},\
  }\bibfield  {title} {\enquote {\bibinfo {title} {Indoor light recycling: a
  new home for organic photovoltaics},}\ }\href@noop {} {\bibfield  {journal}
  {\bibinfo  {journal} {J. Mater. Chem. C}\ }\textbf {\bibinfo {volume} {4}},\
  \bibinfo {pages} {10367--10370} (\bibinfo {year} {2016})}\BibitemShut
  {NoStop}%
\bibitem [{\citenamefont {Davy}\ \emph {et~al.}(2017)\citenamefont {Davy},
  \citenamefont {Sezen-Edmonds}, \citenamefont {Gao}, \citenamefont {Lin},
  \citenamefont {Liu}, \citenamefont {Yao}, \citenamefont {Kahn},\ and\
  \citenamefont {Loo}}]{davy2017}%
  \BibitemOpen
  \bibfield  {author} {\bibinfo {author} {\bibfnamefont {N.~C.}\ \bibnamefont
  {Davy}}, \bibinfo {author} {\bibfnamefont {M.}~\bibnamefont {Sezen-Edmonds}},
  \bibinfo {author} {\bibfnamefont {J.}~\bibnamefont {Gao}}, \bibinfo {author}
  {\bibfnamefont {X.}~\bibnamefont {Lin}}, \bibinfo {author} {\bibfnamefont
  {A.}~\bibnamefont {Liu}}, \bibinfo {author} {\bibfnamefont {N.}~\bibnamefont
  {Yao}}, \bibinfo {author} {\bibfnamefont {A.}~\bibnamefont {Kahn}}, \ and\
  \bibinfo {author} {\bibfnamefont {Y.-L.}\ \bibnamefont {Loo}},\ }\bibfield
  {title} {\enquote {\bibinfo {title} {Pairing of near-ultraviolet solar cells
  with electrochromic windows for smart management of the solar spectrum},}\
  }\href@noop {} {\bibfield  {journal} {\bibinfo  {journal} {Nat. Energy}\
  }\textbf {\bibinfo {volume} {2}},\ \bibinfo {pages} {1--11} (\bibinfo {year}
  {2017})}\BibitemShut {NoStop}%
\bibitem [{\citenamefont {Fu}\ \emph {et~al.}(2021)\citenamefont {Fu},
  \citenamefont {Zhang}, \citenamefont {Zhang}, \citenamefont {Li},
  \citenamefont {Zhou},\ and\ \citenamefont {Zhang}}]{fu2021}%
  \BibitemOpen
  \bibfield  {author} {\bibinfo {author} {\bibfnamefont {Z.}~\bibnamefont
  {Fu}}, \bibinfo {author} {\bibfnamefont {X.}~\bibnamefont {Zhang}}, \bibinfo
  {author} {\bibfnamefont {H.}~\bibnamefont {Zhang}}, \bibinfo {author}
  {\bibfnamefont {Y.}~\bibnamefont {Li}}, \bibinfo {author} {\bibfnamefont
  {H.}~\bibnamefont {Zhou}}, \ and\ \bibinfo {author} {\bibfnamefont
  {Y.}~\bibnamefont {Zhang}},\ }\bibfield  {title} {\enquote {\bibinfo {title}
  {On the understandings of dielectric constant and its impacts on the
  photovoltaic efficiency in organic solar cells},}\ }\href@noop {} {\bibfield
  {journal} {\bibinfo  {journal} {Chinese J. Chem.}\ }\textbf {\bibinfo
  {volume} {39}},\ \bibinfo {pages} {381--390} (\bibinfo {year}
  {2021})}\BibitemShut {NoStop}%
\bibitem [{\citenamefont {Armin}\ \emph {et~al.}(2014)\citenamefont {Armin},
  \citenamefont {Kassal}, \citenamefont {Shaw}, \citenamefont {Hambsch},
  \citenamefont {Stolterfoht}, \citenamefont {Lyons}, \citenamefont {Li},
  \citenamefont {Shi}, \citenamefont {Burn},\ and\ \citenamefont
  {Meredith}}]{armin2014}%
  \BibitemOpen
  \bibfield  {author} {\bibinfo {author} {\bibfnamefont {A.}~\bibnamefont
  {Armin}}, \bibinfo {author} {\bibfnamefont {I.}~\bibnamefont {Kassal}},
  \bibinfo {author} {\bibfnamefont {P.~E.}\ \bibnamefont {Shaw}}, \bibinfo
  {author} {\bibfnamefont {M.}~\bibnamefont {Hambsch}}, \bibinfo {author}
  {\bibfnamefont {M.}~\bibnamefont {Stolterfoht}}, \bibinfo {author}
  {\bibfnamefont {D.~M.}\ \bibnamefont {Lyons}}, \bibinfo {author}
  {\bibfnamefont {J.}~\bibnamefont {Li}}, \bibinfo {author} {\bibfnamefont
  {Z.}~\bibnamefont {Shi}}, \bibinfo {author} {\bibfnamefont {P.~L.}\
  \bibnamefont {Burn}}, \ and\ \bibinfo {author} {\bibfnamefont
  {P.}~\bibnamefont {Meredith}},\ }\bibfield  {title} {\enquote {\bibinfo
  {title} {Spectral dependence of the internal quantum efficiency of organic
  solar cells: Effect of charge generation pathways},}\ }\href@noop {}
  {\bibfield  {journal} {\bibinfo  {journal} {J. Am. Chem. Soc.}\ }\textbf
  {\bibinfo {volume} {136}},\ \bibinfo {pages} {11465--11472} (\bibinfo {year}
  {2014})}\BibitemShut {NoStop}%
\bibitem [{\citenamefont {Stoltzfus}\ \emph {et~al.}(2016)\citenamefont
  {Stoltzfus}, \citenamefont {Donaghey}, \citenamefont {Armin}, \citenamefont
  {Shaw}, \citenamefont {Burn},\ and\ \citenamefont
  {Meredith}}]{stoltzfus2016}%
  \BibitemOpen
  \bibfield  {author} {\bibinfo {author} {\bibfnamefont {D.~M.}\ \bibnamefont
  {Stoltzfus}}, \bibinfo {author} {\bibfnamefont {J.~E.}\ \bibnamefont
  {Donaghey}}, \bibinfo {author} {\bibfnamefont {A.}~\bibnamefont {Armin}},
  \bibinfo {author} {\bibfnamefont {P.~E.}\ \bibnamefont {Shaw}}, \bibinfo
  {author} {\bibfnamefont {P.~L.}\ \bibnamefont {Burn}}, \ and\ \bibinfo
  {author} {\bibfnamefont {P.}~\bibnamefont {Meredith}},\ }\bibfield  {title}
  {\enquote {\bibinfo {title} {Charge generation pathways in organic solar
  cells: assessing the contribution from the electron acceptor},}\ }\href@noop
  {} {\bibfield  {journal} {\bibinfo  {journal} {Chem. Rev.}\ }\textbf
  {\bibinfo {volume} {116}},\ \bibinfo {pages} {12920--12955} (\bibinfo {year}
  {2016})}\BibitemShut {NoStop}%
\bibitem [{\citenamefont {Deibel}\ and\ \citenamefont
  {Dyakonov}(2010)}]{deibel2010}%
  \BibitemOpen
  \bibfield  {author} {\bibinfo {author} {\bibfnamefont {C.}~\bibnamefont
  {Deibel}}\ and\ \bibinfo {author} {\bibfnamefont {V.}~\bibnamefont
  {Dyakonov}},\ }\bibfield  {title} {\enquote {\bibinfo {title}
  {Polymer--fullerene bulk heterojunction solar cells},}\ }\href@noop {}
  {\bibfield  {journal} {\bibinfo  {journal} {Rep. Prog. Phys.}\ }\textbf
  {\bibinfo {volume} {73}},\ \bibinfo {pages} {096401} (\bibinfo {year}
  {2010})}\BibitemShut {NoStop}%
\bibitem [{\citenamefont {Yuan}\ \emph {et~al.}(2019)\citenamefont {Yuan},
  \citenamefont {Zhang}, \citenamefont {Zhou}, \citenamefont {Zhang},
  \citenamefont {Yip}, \citenamefont {Lau}, \citenamefont {Lu}, \citenamefont
  {Zhu}, \citenamefont {Peng}, \citenamefont {Johnson}, \citenamefont
  {Leclerc}, \citenamefont {Cao}, \citenamefont {Ulanski}, \citenamefont {Li},\
  and\ \citenamefont {Zou}}]{yuan2019}%
  \BibitemOpen
  \bibfield  {author} {\bibinfo {author} {\bibfnamefont {J.}~\bibnamefont
  {Yuan}}, \bibinfo {author} {\bibfnamefont {Y.}~\bibnamefont {Zhang}},
  \bibinfo {author} {\bibfnamefont {L.}~\bibnamefont {Zhou}}, \bibinfo {author}
  {\bibfnamefont {G.}~\bibnamefont {Zhang}}, \bibinfo {author} {\bibfnamefont
  {H.-L.}\ \bibnamefont {Yip}}, \bibinfo {author} {\bibfnamefont {T.-K.}\
  \bibnamefont {Lau}}, \bibinfo {author} {\bibfnamefont {X.}~\bibnamefont
  {Lu}}, \bibinfo {author} {\bibfnamefont {C.}~\bibnamefont {Zhu}}, \bibinfo
  {author} {\bibfnamefont {H.}~\bibnamefont {Peng}}, \bibinfo {author}
  {\bibfnamefont {P.~A.}\ \bibnamefont {Johnson}}, \bibinfo {author}
  {\bibfnamefont {M.}~\bibnamefont {Leclerc}}, \bibinfo {author} {\bibfnamefont
  {Y.}~\bibnamefont {Cao}}, \bibinfo {author} {\bibfnamefont {J.}~\bibnamefont
  {Ulanski}}, \bibinfo {author} {\bibfnamefont {Y.}~\bibnamefont {Li}}, \ and\
  \bibinfo {author} {\bibfnamefont {Y.}~\bibnamefont {Zou}},\ }\bibfield
  {title} {\enquote {\bibinfo {title} {Single-junction organic solar cell with
  over 15\% efficiency using fused-ring acceptor with electron-deficient
  core},}\ }\href@noop {} {\bibfield  {journal} {\bibinfo  {journal} {Joule}\
  }\textbf {\bibinfo {volume} {3}},\ \bibinfo {pages} {1140--1151} (\bibinfo
  {year} {2019})}\BibitemShut {NoStop}%
\bibitem [{\citenamefont {Cui}\ \emph {et~al.}(2020)\citenamefont {Cui},
  \citenamefont {Yao}, \citenamefont {Zhang}, \citenamefont {Xian},
  \citenamefont {Zhang}, \citenamefont {Hong}, \citenamefont {Wang},
  \citenamefont {Xu}, \citenamefont {Ma}, \citenamefont {An} \emph
  {et~al.}}]{cui2020}%
  \BibitemOpen
  \bibfield  {author} {\bibinfo {author} {\bibfnamefont {Y.}~\bibnamefont
  {Cui}}, \bibinfo {author} {\bibfnamefont {H.}~\bibnamefont {Yao}}, \bibinfo
  {author} {\bibfnamefont {J.}~\bibnamefont {Zhang}}, \bibinfo {author}
  {\bibfnamefont {K.}~\bibnamefont {Xian}}, \bibinfo {author} {\bibfnamefont
  {T.}~\bibnamefont {Zhang}}, \bibinfo {author} {\bibfnamefont
  {L.}~\bibnamefont {Hong}}, \bibinfo {author} {\bibfnamefont {Y.}~\bibnamefont
  {Wang}}, \bibinfo {author} {\bibfnamefont {Y.}~\bibnamefont {Xu}}, \bibinfo
  {author} {\bibfnamefont {K.}~\bibnamefont {Ma}}, \bibinfo {author}
  {\bibfnamefont {C.}~\bibnamefont {An}},  \emph {et~al.},\ }\bibfield  {title}
  {\enquote {\bibinfo {title} {Single-junction organic photovoltaic cells with
  approaching 18\% efficiency},}\ }\href@noop {} {\bibfield  {journal}
  {\bibinfo  {journal} {Adv. Mater.}\ }\textbf {\bibinfo {volume} {32}},\
  \bibinfo {pages} {1908205} (\bibinfo {year} {2020})}\BibitemShut {NoStop}%
\bibitem [{\citenamefont {Qin}\ \emph {et~al.}(2020)\citenamefont {Qin},
  \citenamefont {Zhang}, \citenamefont {Xiao}, \citenamefont {Chen},
  \citenamefont {Sun}, \citenamefont {Zang}, \citenamefont {Yi}, \citenamefont
  {Yuan}, \citenamefont {Jin}, \citenamefont {Hao} \emph {et~al.}}]{qin2020}%
  \BibitemOpen
  \bibfield  {author} {\bibinfo {author} {\bibfnamefont {J.}~\bibnamefont
  {Qin}}, \bibinfo {author} {\bibfnamefont {L.}~\bibnamefont {Zhang}}, \bibinfo
  {author} {\bibfnamefont {Z.}~\bibnamefont {Xiao}}, \bibinfo {author}
  {\bibfnamefont {S.}~\bibnamefont {Chen}}, \bibinfo {author} {\bibfnamefont
  {K.}~\bibnamefont {Sun}}, \bibinfo {author} {\bibfnamefont {Z.}~\bibnamefont
  {Zang}}, \bibinfo {author} {\bibfnamefont {C.}~\bibnamefont {Yi}}, \bibinfo
  {author} {\bibfnamefont {Y.}~\bibnamefont {Yuan}}, \bibinfo {author}
  {\bibfnamefont {Z.}~\bibnamefont {Jin}}, \bibinfo {author} {\bibfnamefont
  {F.}~\bibnamefont {Hao}},  \emph {et~al.},\ }\bibfield  {title} {\enquote
  {\bibinfo {title} {Over 16\% efficiency from thick-film organic solar
  cells},}\ }\href@noop {} {\bibfield  {journal} {\bibinfo  {journal} {Sci.
  Bull.}\ }\textbf {\bibinfo {volume} {65}},\ \bibinfo {pages} {1979--82}
  (\bibinfo {year} {2020})}\BibitemShut {NoStop}%
\bibitem [{\citenamefont {Liu}\ \emph {et~al.}(2020{\natexlab{a}})\citenamefont
  {Liu}, \citenamefont {Yuan}, \citenamefont {Deng}, \citenamefont {Luo},
  \citenamefont {Xie}, \citenamefont {Liang}, \citenamefont {Zou},
  \citenamefont {He}, \citenamefont {Wu},\ and\ \citenamefont {Cao}}]{liu2020}%
  \BibitemOpen
  \bibfield  {author} {\bibinfo {author} {\bibfnamefont {S.}~\bibnamefont
  {Liu}}, \bibinfo {author} {\bibfnamefont {J.}~\bibnamefont {Yuan}}, \bibinfo
  {author} {\bibfnamefont {W.}~\bibnamefont {Deng}}, \bibinfo {author}
  {\bibfnamefont {M.}~\bibnamefont {Luo}}, \bibinfo {author} {\bibfnamefont
  {Y.}~\bibnamefont {Xie}}, \bibinfo {author} {\bibfnamefont {Q.}~\bibnamefont
  {Liang}}, \bibinfo {author} {\bibfnamefont {Y.}~\bibnamefont {Zou}}, \bibinfo
  {author} {\bibfnamefont {Z.}~\bibnamefont {He}}, \bibinfo {author}
  {\bibfnamefont {H.}~\bibnamefont {Wu}}, \ and\ \bibinfo {author}
  {\bibfnamefont {Y.}~\bibnamefont {Cao}},\ }\bibfield  {title} {\enquote
  {\bibinfo {title} {High-efficiency organic solar cells with low non-radiative
  recombination loss and low energetic disorder},}\ }\href@noop {} {\bibfield
  {journal} {\bibinfo  {journal} {Nat. Photonics}\ }\textbf {\bibinfo {volume}
  {14}},\ \bibinfo {pages} {300--305} (\bibinfo {year}
  {2020}{\natexlab{a}})}\BibitemShut {NoStop}%
\bibitem [{\citenamefont {Liu}\ \emph {et~al.}(2020{\natexlab{b}})\citenamefont
  {Liu}, \citenamefont {Jiang}, \citenamefont {Jin}, \citenamefont {Qin},
  \citenamefont {Xu}, \citenamefont {Li}, \citenamefont {Xiong}, \citenamefont
  {Liu}, \citenamefont {Xiao}, \citenamefont {Sun}, \citenamefont {Yang},\ and\
  \citenamefont {Zhang}}]{liu22020}%
  \BibitemOpen
  \bibfield  {author} {\bibinfo {author} {\bibfnamefont {Q.}~\bibnamefont
  {Liu}}, \bibinfo {author} {\bibfnamefont {Y.}~\bibnamefont {Jiang}}, \bibinfo
  {author} {\bibfnamefont {K.}~\bibnamefont {Jin}}, \bibinfo {author}
  {\bibfnamefont {J.}~\bibnamefont {Qin}}, \bibinfo {author} {\bibfnamefont
  {J.}~\bibnamefont {Xu}}, \bibinfo {author} {\bibfnamefont {W.}~\bibnamefont
  {Li}}, \bibinfo {author} {\bibfnamefont {J.}~\bibnamefont {Xiong}}, \bibinfo
  {author} {\bibfnamefont {J.}~\bibnamefont {Liu}}, \bibinfo {author}
  {\bibfnamefont {Z.}~\bibnamefont {Xiao}}, \bibinfo {author} {\bibfnamefont
  {K.}~\bibnamefont {Sun}}, \bibinfo {author} {\bibfnamefont {S.}~\bibnamefont
  {Yang}}, \ and\ \bibinfo {author} {\bibfnamefont {X.}~\bibnamefont {Zhang}},\
  }\bibfield  {title} {\enquote {\bibinfo {title} {18\% efficiency organic
  solar cells},}\ }\href@noop {} {\bibfield  {journal} {\bibinfo  {journal}
  {Sci. Bull.}\ }\textbf {\bibinfo {volume} {65}},\ \bibinfo {pages} {272--275}
  (\bibinfo {year} {2020}{\natexlab{b}})}\BibitemShut {NoStop}%
\bibitem [{SUP()}]{SUPP}%
  \BibitemOpen
  \href@noop {} {\ }\bibinfo {note} {See Supplemental Material at
  [\textbf{***}] for further information on chemical definitions, thin film
  deposition, spotsize measurements, and error analysis.}\BibitemShut {Stop}%
\bibitem [{\citenamefont {Chen}\ \emph {et~al.}(2017)\citenamefont {Chen},
  \citenamefont {Liu}, \citenamefont {Zhang}, \citenamefont {Chow},
  \citenamefont {Wang}, \citenamefont {Zhang}, \citenamefont {Ma},\ and\
  \citenamefont {Yan}}]{chen2017}%
  \BibitemOpen
  \bibfield  {author} {\bibinfo {author} {\bibfnamefont {S.}~\bibnamefont
  {Chen}}, \bibinfo {author} {\bibfnamefont {Y.}~\bibnamefont {Liu}}, \bibinfo
  {author} {\bibfnamefont {L.}~\bibnamefont {Zhang}}, \bibinfo {author}
  {\bibfnamefont {P.~C.}\ \bibnamefont {Chow}}, \bibinfo {author}
  {\bibfnamefont {Z.}~\bibnamefont {Wang}}, \bibinfo {author} {\bibfnamefont
  {G.}~\bibnamefont {Zhang}}, \bibinfo {author} {\bibfnamefont
  {W.}~\bibnamefont {Ma}}, \ and\ \bibinfo {author} {\bibfnamefont
  {H.}~\bibnamefont {Yan}},\ }\bibfield  {title} {\enquote {\bibinfo {title} {A
  wide-bandgap donor polymer for highly efficient non-fullerene organic solar
  cells with a small voltage loss},}\ }\href@noop {} {\bibfield  {journal}
  {\bibinfo  {journal} {J. Am. Chem. Soc.}\ }\textbf {\bibinfo {volume}
  {139}},\ \bibinfo {pages} {6298--6301} (\bibinfo {year} {2017})}\BibitemShut
  {NoStop}%
\bibitem [{\citenamefont {Xiao}\ \emph {et~al.}(2017)\citenamefont {Xiao},
  \citenamefont {Jia}, \citenamefont {Li}, \citenamefont {Wang}, \citenamefont
  {Geng}, \citenamefont {Liu}, \citenamefont {Chen}, \citenamefont {Yang},
  \citenamefont {Russell},\ and\ \citenamefont {Ding}}]{xiao2017}%
  \BibitemOpen
  \bibfield  {author} {\bibinfo {author} {\bibfnamefont {Z.}~\bibnamefont
  {Xiao}}, \bibinfo {author} {\bibfnamefont {X.}~\bibnamefont {Jia}}, \bibinfo
  {author} {\bibfnamefont {D.}~\bibnamefont {Li}}, \bibinfo {author}
  {\bibfnamefont {S.}~\bibnamefont {Wang}}, \bibinfo {author} {\bibfnamefont
  {X.}~\bibnamefont {Geng}}, \bibinfo {author} {\bibfnamefont {F.}~\bibnamefont
  {Liu}}, \bibinfo {author} {\bibfnamefont {J.}~\bibnamefont {Chen}}, \bibinfo
  {author} {\bibfnamefont {S.}~\bibnamefont {Yang}}, \bibinfo {author}
  {\bibfnamefont {T.~P.}\ \bibnamefont {Russell}}, \ and\ \bibinfo {author}
  {\bibfnamefont {L.}~\bibnamefont {Ding}},\ }\bibfield  {title} {\enquote
  {\bibinfo {title} {26 m{A}/$\text{cm}^2$ {J}sc from organic solar cells with
  a low-bandgap nonfullerene acceptor},}\ }\href@noop {} {\bibfield  {journal}
  {\bibinfo  {journal} {Sci. Bull.}\ }\textbf {\bibinfo {volume} {62}},\
  \bibinfo {pages} {1494--1496} (\bibinfo {year} {2017})}\BibitemShut {NoStop}%
\bibitem [{\citenamefont {Wang}\ \emph {et~al.}(2021)\citenamefont {Wang},
  \citenamefont {Chen}, \citenamefont {Yuan}, \citenamefont {Wei},
  \citenamefont {Li}, \citenamefont {Jiang}, \citenamefont {Huang},
  \citenamefont {Li}, \citenamefont {Li},\ and\ \citenamefont
  {Zou}}]{wang2021}%
  \BibitemOpen
  \bibfield  {author} {\bibinfo {author} {\bibfnamefont {X.}~\bibnamefont
  {Wang}}, \bibinfo {author} {\bibfnamefont {H.}~\bibnamefont {Chen}}, \bibinfo
  {author} {\bibfnamefont {J.}~\bibnamefont {Yuan}}, \bibinfo {author}
  {\bibfnamefont {Q.}~\bibnamefont {Wei}}, \bibinfo {author} {\bibfnamefont
  {J.}~\bibnamefont {Li}}, \bibinfo {author} {\bibfnamefont {L.}~\bibnamefont
  {Jiang}}, \bibinfo {author} {\bibfnamefont {J.}~\bibnamefont {Huang}},
  \bibinfo {author} {\bibfnamefont {Y.}~\bibnamefont {Li}}, \bibinfo {author}
  {\bibfnamefont {Y.}~\bibnamefont {Li}}, \ and\ \bibinfo {author}
  {\bibfnamefont {Y.}~\bibnamefont {Zou}},\ }\bibfield  {title} {\enquote
  {\bibinfo {title} {Precise fluorination of polymeric donors towards efficient
  non-fullerene organic solar cells with enhanced open circuit voltage, short
  circuit current and fill factor},}\ }\href@noop {} {\bibfield  {journal}
  {\bibinfo  {journal} {J. Mater. Chem. A}\ } (\bibinfo {year}
  {2021})}\BibitemShut {NoStop}%
\bibitem [{\citenamefont {Sajjad}\ \emph {et~al.}(2020)\citenamefont {Sajjad},
  \citenamefont {Ruseckas}, \citenamefont {Jagadamma}, \citenamefont {Zhang},\
  and\ \citenamefont {Samuel}}]{sajjad2020}%
  \BibitemOpen
  \bibfield  {author} {\bibinfo {author} {\bibfnamefont {M.~T.}\ \bibnamefont
  {Sajjad}}, \bibinfo {author} {\bibfnamefont {A.}~\bibnamefont {Ruseckas}},
  \bibinfo {author} {\bibfnamefont {L.~K.}\ \bibnamefont {Jagadamma}}, \bibinfo
  {author} {\bibfnamefont {Y.}~\bibnamefont {Zhang}}, \ and\ \bibinfo {author}
  {\bibfnamefont {I.~D.}\ \bibnamefont {Samuel}},\ }\bibfield  {title}
  {\enquote {\bibinfo {title} {Long-range exciton diffusion in non-fullerene
  acceptors and coarse bulk heterojunctions enable highly efficient organic
  photovoltaics},}\ }\href@noop {} {\bibfield  {journal} {\bibinfo  {journal}
  {J. Mater. Chem. A}\ } (\bibinfo {year} {2020})}\BibitemShut {NoStop}%
\bibitem [{\citenamefont {Classen}\ \emph {et~al.}(2020)\citenamefont
  {Classen}, \citenamefont {Chochos}, \citenamefont {L{\"u}er}, \citenamefont
  {Gregoriou}, \citenamefont {Wortmann}, \citenamefont {Osvet}, \citenamefont
  {Forberich}, \citenamefont {McCulloch}, \citenamefont {Heum{\"u}ller},\ and\
  \citenamefont {Brabec}}]{classen2020}%
  \BibitemOpen
  \bibfield  {author} {\bibinfo {author} {\bibfnamefont {A.}~\bibnamefont
  {Classen}}, \bibinfo {author} {\bibfnamefont {C.~L.}\ \bibnamefont
  {Chochos}}, \bibinfo {author} {\bibfnamefont {L.}~\bibnamefont {L{\"u}er}},
  \bibinfo {author} {\bibfnamefont {V.~G.}\ \bibnamefont {Gregoriou}}, \bibinfo
  {author} {\bibfnamefont {J.}~\bibnamefont {Wortmann}}, \bibinfo {author}
  {\bibfnamefont {A.}~\bibnamefont {Osvet}}, \bibinfo {author} {\bibfnamefont
  {K.}~\bibnamefont {Forberich}}, \bibinfo {author} {\bibfnamefont
  {I.}~\bibnamefont {McCulloch}}, \bibinfo {author} {\bibfnamefont
  {T.}~\bibnamefont {Heum{\"u}ller}}, \ and\ \bibinfo {author} {\bibfnamefont
  {C.~J.}\ \bibnamefont {Brabec}},\ }\bibfield  {title} {\enquote {\bibinfo
  {title} {The role of exciton lifetime for charge generation in organic solar
  cells at negligible energy-level offsets},}\ }\href@noop {} {\bibfield
  {journal} {\bibinfo  {journal} {Nat. Energy}\ }\textbf {\bibinfo {volume}
  {5}},\ \bibinfo {pages} {711--719} (\bibinfo {year} {2020})}\BibitemShut
  {NoStop}%
\bibitem [{\citenamefont {Karuthedath}\ \emph {et~al.}(2021)\citenamefont
  {Karuthedath}, \citenamefont {Gorenflot}, \citenamefont {Firdaus},
  \citenamefont {Chaturvedi}, \citenamefont {De~Castro}, \citenamefont
  {Harrison}, \citenamefont {Khan}, \citenamefont {Markina}, \citenamefont
  {Balawi}, \citenamefont {Pe{\~n}a} \emph {et~al.}}]{karuthedath2021}%
  \BibitemOpen
  \bibfield  {author} {\bibinfo {author} {\bibfnamefont {S.}~\bibnamefont
  {Karuthedath}}, \bibinfo {author} {\bibfnamefont {J.}~\bibnamefont
  {Gorenflot}}, \bibinfo {author} {\bibfnamefont {Y.}~\bibnamefont {Firdaus}},
  \bibinfo {author} {\bibfnamefont {N.}~\bibnamefont {Chaturvedi}}, \bibinfo
  {author} {\bibfnamefont {C.~S.}\ \bibnamefont {De~Castro}}, \bibinfo {author}
  {\bibfnamefont {G.~T.}\ \bibnamefont {Harrison}}, \bibinfo {author}
  {\bibfnamefont {J.~I.}\ \bibnamefont {Khan}}, \bibinfo {author}
  {\bibfnamefont {A.}~\bibnamefont {Markina}}, \bibinfo {author} {\bibfnamefont
  {A.~H.}\ \bibnamefont {Balawi}}, \bibinfo {author} {\bibfnamefont {T.~A.~D.}\
  \bibnamefont {Pe{\~n}a}},  \emph {et~al.},\ }\bibfield  {title} {\enquote
  {\bibinfo {title} {Intrinsic efficiency limits in low-bandgap non-fullerene
  acceptor organic solar cells},}\ }\href@noop {} {\bibfield  {journal}
  {\bibinfo  {journal} {Nat. Mater.}\ }\textbf {\bibinfo {volume} {20}},\
  \bibinfo {pages} {378--384} (\bibinfo {year} {2021})}\BibitemShut {NoStop}%
\bibitem [{\citenamefont {F{\"o}rster}(1948)}]{forster1948}%
  \BibitemOpen
  \bibfield  {author} {\bibinfo {author} {\bibfnamefont {T.}~\bibnamefont
  {F{\"o}rster}},\ }\bibfield  {title} {\enquote {\bibinfo {title} {Energy
  transfer and fluorescence between molecules},}\ }\href@noop {} {\bibfield
  {journal} {\bibinfo  {journal} {Ann. Phys.}\ }\textbf {\bibinfo {volume}
  {437}},\ \bibinfo {pages} {55--75} (\bibinfo {year} {1948})}\BibitemShut
  {NoStop}%
\bibitem [{\citenamefont {Ward}\ \emph {et~al.}(2012)\citenamefont {Ward},
  \citenamefont {Ruseckas},\ and\ \citenamefont {Samuel}}]{ward2012}%
  \BibitemOpen
  \bibfield  {author} {\bibinfo {author} {\bibfnamefont {A.~J.}\ \bibnamefont
  {Ward}}, \bibinfo {author} {\bibfnamefont {A.}~\bibnamefont {Ruseckas}}, \
  and\ \bibinfo {author} {\bibfnamefont {I.~D.}\ \bibnamefont {Samuel}},\
  }\bibfield  {title} {\enquote {\bibinfo {title} {A shift from diffusion
  assisted to energy transfer controlled fluorescence quenching in
  polymer--fullerene photovoltaic blends},}\ }\href@noop {} {\bibfield
  {journal} {\bibinfo  {journal} {J. Phys. Chem. C}\ }\textbf {\bibinfo
  {volume} {116}},\ \bibinfo {pages} {23931--23937} (\bibinfo {year}
  {2012})}\BibitemShut {NoStop}%
\bibitem [{\citenamefont {Mikhnenko}\ \emph {et~al.}(2012)\citenamefont
  {Mikhnenko}, \citenamefont {Azimi}, \citenamefont {Scharber}, \citenamefont
  {Morana}, \citenamefont {Blom},\ and\ \citenamefont {Loi}}]{mikhnenko2012}%
  \BibitemOpen
  \bibfield  {author} {\bibinfo {author} {\bibfnamefont {O.~V.}\ \bibnamefont
  {Mikhnenko}}, \bibinfo {author} {\bibfnamefont {H.}~\bibnamefont {Azimi}},
  \bibinfo {author} {\bibfnamefont {M.}~\bibnamefont {Scharber}}, \bibinfo
  {author} {\bibfnamefont {M.}~\bibnamefont {Morana}}, \bibinfo {author}
  {\bibfnamefont {P.~W.}\ \bibnamefont {Blom}}, \ and\ \bibinfo {author}
  {\bibfnamefont {M.~A.}\ \bibnamefont {Loi}},\ }\bibfield  {title} {\enquote
  {\bibinfo {title} {Exciton diffusion length in narrow bandgap polymers},}\
  }\href@noop {} {\bibfield  {journal} {\bibinfo  {journal} {Energy Environ.
  Sci.}\ }\textbf {\bibinfo {volume} {5}},\ \bibinfo {pages} {6960--6965}
  (\bibinfo {year} {2012})}\BibitemShut {NoStop}%
\bibitem [{\citenamefont {Wang}\ \emph {et~al.}(2011)\citenamefont {Wang},
  \citenamefont {Wang}, \citenamefont {Gao}, \citenamefont {Wang},
  \citenamefont {Yang}, \citenamefont {Du}, \citenamefont {Chen}, \citenamefont
  {Song},\ and\ \citenamefont {Sun}}]{wang2011}%
  \BibitemOpen
  \bibfield  {author} {\bibinfo {author} {\bibfnamefont {H.}~\bibnamefont
  {Wang}}, \bibinfo {author} {\bibfnamefont {H.-Y.}\ \bibnamefont {Wang}},
  \bibinfo {author} {\bibfnamefont {B.-R.}\ \bibnamefont {Gao}}, \bibinfo
  {author} {\bibfnamefont {L.}~\bibnamefont {Wang}}, \bibinfo {author}
  {\bibfnamefont {Z.-Y.}\ \bibnamefont {Yang}}, \bibinfo {author}
  {\bibfnamefont {X.-B.}\ \bibnamefont {Du}}, \bibinfo {author} {\bibfnamefont
  {Q.-D.}\ \bibnamefont {Chen}}, \bibinfo {author} {\bibfnamefont {J.-F.}\
  \bibnamefont {Song}}, \ and\ \bibinfo {author} {\bibfnamefont {H.-B.}\
  \bibnamefont {Sun}},\ }\bibfield  {title} {\enquote {\bibinfo {title}
  {Exciton diffusion and charge transfer dynamics in nano phase-separated
  {P3HT}/{PCBM} blend films},}\ }\href@noop {} {\bibfield  {journal} {\bibinfo
  {journal} {Nanoscale}\ }\textbf {\bibinfo {volume} {3}},\ \bibinfo {pages}
  {2280--2285} (\bibinfo {year} {2011})}\BibitemShut {NoStop}%
\bibitem [{\citenamefont {Theander}\ \emph {et~al.}(2000)\citenamefont
  {Theander}, \citenamefont {Yartsev}, \citenamefont {Zigmantas}, \citenamefont
  {Sundstr{\"o}m}, \citenamefont {Mammo}, \citenamefont {Andersson},\ and\
  \citenamefont {Ingan{\"a}s}}]{theander2000}%
  \BibitemOpen
  \bibfield  {author} {\bibinfo {author} {\bibfnamefont {M.}~\bibnamefont
  {Theander}}, \bibinfo {author} {\bibfnamefont {A.}~\bibnamefont {Yartsev}},
  \bibinfo {author} {\bibfnamefont {D.}~\bibnamefont {Zigmantas}}, \bibinfo
  {author} {\bibfnamefont {V.}~\bibnamefont {Sundstr{\"o}m}}, \bibinfo {author}
  {\bibfnamefont {W.}~\bibnamefont {Mammo}}, \bibinfo {author} {\bibfnamefont
  {M.~R.}\ \bibnamefont {Andersson}}, \ and\ \bibinfo {author} {\bibfnamefont
  {O.}~\bibnamefont {Ingan{\"a}s}},\ }\bibfield  {title} {\enquote {\bibinfo
  {title} {Photoluminescence quenching at a polythiophene/${C}_{60}$
  heterojunction},}\ }\href@noop {} {\bibfield  {journal} {\bibinfo  {journal}
  {Phys. Rev. B}\ }\textbf {\bibinfo {volume} {61}},\ \bibinfo {pages} {12957}
  (\bibinfo {year} {2000})}\BibitemShut {NoStop}%
\bibitem [{\citenamefont {Luhman}\ and\ \citenamefont
  {Holmes}(2011)}]{luhman2011}%
  \BibitemOpen
  \bibfield  {author} {\bibinfo {author} {\bibfnamefont {W.~A.}\ \bibnamefont
  {Luhman}}\ and\ \bibinfo {author} {\bibfnamefont {R.~J.}\ \bibnamefont
  {Holmes}},\ }\bibfield  {title} {\enquote {\bibinfo {title} {Investigation of
  energy transfer in organic photovoltaic cells and impact on exciton diffusion
  length measurements},}\ }\href@noop {} {\bibfield  {journal} {\bibinfo
  {journal} {Adv. Funct. Mater.}\ }\textbf {\bibinfo {volume} {21}},\ \bibinfo
  {pages} {764--771} (\bibinfo {year} {2011})}\BibitemShut {NoStop}%
\bibitem [{\citenamefont {Mikhnenko}\ \emph {et~al.}(2008)\citenamefont
  {Mikhnenko}, \citenamefont {Cordella}, \citenamefont {Sieval}, \citenamefont
  {Hummelen}, \citenamefont {Blom},\ and\ \citenamefont {Loi}}]{mikhnenko2008}%
  \BibitemOpen
  \bibfield  {author} {\bibinfo {author} {\bibfnamefont {O.}~\bibnamefont
  {Mikhnenko}}, \bibinfo {author} {\bibfnamefont {F.}~\bibnamefont {Cordella}},
  \bibinfo {author} {\bibfnamefont {A.}~\bibnamefont {Sieval}}, \bibinfo
  {author} {\bibfnamefont {J.}~\bibnamefont {Hummelen}}, \bibinfo {author}
  {\bibfnamefont {P.}~\bibnamefont {Blom}}, \ and\ \bibinfo {author}
  {\bibfnamefont {M.}~\bibnamefont {Loi}},\ }\bibfield  {title} {\enquote
  {\bibinfo {title} {Temperature dependence of exciton diffusion in conjugated
  polymers},}\ }\href@noop {} {\bibfield  {journal} {\bibinfo  {journal} {J.
  Phys. Chem. B}\ }\textbf {\bibinfo {volume} {112}},\ \bibinfo {pages}
  {11601--11604} (\bibinfo {year} {2008})}\BibitemShut {NoStop}%
\bibitem [{\citenamefont {Haugeneder}\ \emph {et~al.}(1999)\citenamefont
  {Haugeneder}, \citenamefont {Neges}, \citenamefont {Kallinger}, \citenamefont
  {Spirkl}, \citenamefont {Lemmer}, \citenamefont {Feldmann}, \citenamefont
  {Scherf}, \citenamefont {Harth}, \citenamefont {G{\"u}gel},\ and\
  \citenamefont {M{\"u}llen}}]{haugeneder1999}%
  \BibitemOpen
  \bibfield  {author} {\bibinfo {author} {\bibfnamefont {A.}~\bibnamefont
  {Haugeneder}}, \bibinfo {author} {\bibfnamefont {M.}~\bibnamefont {Neges}},
  \bibinfo {author} {\bibfnamefont {C.}~\bibnamefont {Kallinger}}, \bibinfo
  {author} {\bibfnamefont {W.}~\bibnamefont {Spirkl}}, \bibinfo {author}
  {\bibfnamefont {U.}~\bibnamefont {Lemmer}}, \bibinfo {author} {\bibfnamefont
  {J.}~\bibnamefont {Feldmann}}, \bibinfo {author} {\bibfnamefont
  {U.}~\bibnamefont {Scherf}}, \bibinfo {author} {\bibfnamefont
  {E.}~\bibnamefont {Harth}}, \bibinfo {author} {\bibfnamefont
  {A.}~\bibnamefont {G{\"u}gel}}, \ and\ \bibinfo {author} {\bibfnamefont
  {K.}~\bibnamefont {M{\"u}llen}},\ }\bibfield  {title} {\enquote {\bibinfo
  {title} {Exciton diffusion and dissociation in conjugated polymer/fullerene
  blends and heterostructures},}\ }\href@noop {} {\bibfield  {journal}
  {\bibinfo  {journal} {Phys. Rev. B}\ }\textbf {\bibinfo {volume} {59}},\
  \bibinfo {pages} {15346} (\bibinfo {year} {1999})}\BibitemShut {NoStop}%
\bibitem [{\citenamefont {Mikhnenko}\ \emph {et~al.}(2009)\citenamefont
  {Mikhnenko}, \citenamefont {Cordella}, \citenamefont {Sieval}, \citenamefont
  {Hummelen}, \citenamefont {Blom},\ and\ \citenamefont {Loi}}]{mikhnenko2009}%
  \BibitemOpen
  \bibfield  {author} {\bibinfo {author} {\bibfnamefont {O.~V.}\ \bibnamefont
  {Mikhnenko}}, \bibinfo {author} {\bibfnamefont {F.}~\bibnamefont {Cordella}},
  \bibinfo {author} {\bibfnamefont {A.~B.}\ \bibnamefont {Sieval}}, \bibinfo
  {author} {\bibfnamefont {J.~C.}\ \bibnamefont {Hummelen}}, \bibinfo {author}
  {\bibfnamefont {P.~W.}\ \bibnamefont {Blom}}, \ and\ \bibinfo {author}
  {\bibfnamefont {M.~A.}\ \bibnamefont {Loi}},\ }\bibfield  {title} {\enquote
  {\bibinfo {title} {Exciton quenching close to polymer- vacuum interface of
  spin-coated films of poly (p-phenylenevinylene) derivative},}\ }\href@noop {}
  {\bibfield  {journal} {\bibinfo  {journal} {J. Phys. Chem. B}\ }\textbf
  {\bibinfo {volume} {113}},\ \bibinfo {pages} {9104--9109} (\bibinfo {year}
  {2009})}\BibitemShut {NoStop}%
\bibitem [{\citenamefont {Scully}\ and\ \citenamefont
  {McGehee}(2006)}]{scully2006}%
  \BibitemOpen
  \bibfield  {author} {\bibinfo {author} {\bibfnamefont {S.~R.}\ \bibnamefont
  {Scully}}\ and\ \bibinfo {author} {\bibfnamefont {M.~D.}\ \bibnamefont
  {McGehee}},\ }\bibfield  {title} {\enquote {\bibinfo {title} {Effects of
  optical interference and energy transfer on exciton diffusion length
  measurements in organic semiconductors},}\ }\href@noop {} {\bibfield
  {journal} {\bibinfo  {journal} {J. App. Phys.}\ }\textbf {\bibinfo {volume}
  {100}},\ \bibinfo {pages} {034907} (\bibinfo {year} {2006})}\BibitemShut
  {NoStop}%
\bibitem [{\citenamefont {Markov}\ \emph {et~al.}(2005)\citenamefont {Markov},
  \citenamefont {Amsterdam}, \citenamefont {Blom}, \citenamefont {Sieval},\
  and\ \citenamefont {Hummelen}}]{markov2005}%
  \BibitemOpen
  \bibfield  {author} {\bibinfo {author} {\bibfnamefont {D.~E.}\ \bibnamefont
  {Markov}}, \bibinfo {author} {\bibfnamefont {E.}~\bibnamefont {Amsterdam}},
  \bibinfo {author} {\bibfnamefont {P.~W.}\ \bibnamefont {Blom}}, \bibinfo
  {author} {\bibfnamefont {A.~B.}\ \bibnamefont {Sieval}}, \ and\ \bibinfo
  {author} {\bibfnamefont {J.~C.}\ \bibnamefont {Hummelen}},\ }\bibfield
  {title} {\enquote {\bibinfo {title} {Accurate measurement of the exciton
  diffusion length in a conjugated polymer using a heterostructure with a
  side-chain cross-linked fullerene layer},}\ }\href@noop {} {\bibfield
  {journal} {\bibinfo  {journal} {J. Phys. Chem. A}\ }\textbf {\bibinfo
  {volume} {109}},\ \bibinfo {pages} {5266--5274} (\bibinfo {year}
  {2005})}\BibitemShut {NoStop}%
\bibitem [{\citenamefont {Shaw}\ \emph {et~al.}(2008)\citenamefont {Shaw},
  \citenamefont {Ruseckas},\ and\ \citenamefont {Samuel}}]{shaw2008}%
  \BibitemOpen
  \bibfield  {author} {\bibinfo {author} {\bibfnamefont {P.~E.}\ \bibnamefont
  {Shaw}}, \bibinfo {author} {\bibfnamefont {A.}~\bibnamefont {Ruseckas}}, \
  and\ \bibinfo {author} {\bibfnamefont {I.~D.}\ \bibnamefont {Samuel}},\
  }\bibfield  {title} {\enquote {\bibinfo {title} {Exciton diffusion
  measurements in poly (3-hexylthiophene)},}\ }\href@noop {} {\bibfield
  {journal} {\bibinfo  {journal} {Adv. Mater.}\ }\textbf {\bibinfo {volume}
  {20}},\ \bibinfo {pages} {3516--3520} (\bibinfo {year} {2008})}\BibitemShut
  {NoStop}%
\bibitem [{\citenamefont {Lewis}\ \emph {et~al.}(2006)\citenamefont {Lewis},
  \citenamefont {Ruseckas}, \citenamefont {Gaudin}, \citenamefont {Webster},
  \citenamefont {Burn},\ and\ \citenamefont {Samuel}}]{lewis2006}%
  \BibitemOpen
  \bibfield  {author} {\bibinfo {author} {\bibfnamefont {A.}~\bibnamefont
  {Lewis}}, \bibinfo {author} {\bibfnamefont {A.}~\bibnamefont {Ruseckas}},
  \bibinfo {author} {\bibfnamefont {O.}~\bibnamefont {Gaudin}}, \bibinfo
  {author} {\bibfnamefont {G.}~\bibnamefont {Webster}}, \bibinfo {author}
  {\bibfnamefont {P.}~\bibnamefont {Burn}}, \ and\ \bibinfo {author}
  {\bibfnamefont {I.}~\bibnamefont {Samuel}},\ }\bibfield  {title} {\enquote
  {\bibinfo {title} {Singlet exciton diffusion in {MEH}-{PPV} films studied by
  exciton--exciton annihilation},}\ }\href@noop {} {\bibfield  {journal}
  {\bibinfo  {journal} {Org. Electron.}\ }\textbf {\bibinfo {volume} {7}},\
  \bibinfo {pages} {452--456} (\bibinfo {year} {2006})}\BibitemShut {NoStop}%
\bibitem [{\citenamefont {Engel}\ \emph {et~al.}(2006)\citenamefont {Engel},
  \citenamefont {Leo},\ and\ \citenamefont {Hoffmann}}]{engel2006}%
  \BibitemOpen
  \bibfield  {author} {\bibinfo {author} {\bibfnamefont {E.}~\bibnamefont
  {Engel}}, \bibinfo {author} {\bibfnamefont {K.}~\bibnamefont {Leo}}, \ and\
  \bibinfo {author} {\bibfnamefont {M.}~\bibnamefont {Hoffmann}},\ }\bibfield
  {title} {\enquote {\bibinfo {title} {Ultrafast relaxation and
  exciton--exciton annihilation in {PTCDA} thin films at high excitation
  densities},}\ }\href@noop {} {\bibfield  {journal} {\bibinfo  {journal}
  {Chem. Phys.}\ }\textbf {\bibinfo {volume} {325}},\ \bibinfo {pages}
  {170--177} (\bibinfo {year} {2006})}\BibitemShut {NoStop}%
\bibitem [{\citenamefont {Cook}\ \emph {et~al.}(2009)\citenamefont {Cook},
  \citenamefont {Furube}, \citenamefont {Katoh},\ and\ \citenamefont
  {Han}}]{cook2009}%
  \BibitemOpen
  \bibfield  {author} {\bibinfo {author} {\bibfnamefont {S.}~\bibnamefont
  {Cook}}, \bibinfo {author} {\bibfnamefont {A.}~\bibnamefont {Furube}},
  \bibinfo {author} {\bibfnamefont {R.}~\bibnamefont {Katoh}}, \ and\ \bibinfo
  {author} {\bibfnamefont {L.}~\bibnamefont {Han}},\ }\bibfield  {title}
  {\enquote {\bibinfo {title} {Estimate of singlet diffusion lengths in {PCBM}
  films by time-resolved emission studies},}\ }\href@noop {} {\bibfield
  {journal} {\bibinfo  {journal} {Chem. Phys. Lett.}\ }\textbf {\bibinfo
  {volume} {478}},\ \bibinfo {pages} {33--36} (\bibinfo {year}
  {2009})}\BibitemShut {NoStop}%
\bibitem [{\citenamefont {Cook}\ \emph {et~al.}(2010)\citenamefont {Cook},
  \citenamefont {Liyuan}, \citenamefont {Furube},\ and\ \citenamefont
  {Katoh}}]{cook2010}%
  \BibitemOpen
  \bibfield  {author} {\bibinfo {author} {\bibfnamefont {S.}~\bibnamefont
  {Cook}}, \bibinfo {author} {\bibfnamefont {H.}~\bibnamefont {Liyuan}},
  \bibinfo {author} {\bibfnamefont {A.}~\bibnamefont {Furube}}, \ and\ \bibinfo
  {author} {\bibfnamefont {R.}~\bibnamefont {Katoh}},\ }\bibfield  {title}
  {\enquote {\bibinfo {title} {Singlet annihilation in films of regioregular
  poly (3-hexylthiophene): Estimates for singlet diffusion lengths and the
  correlation between singlet annihilation rates and spectral relaxation},}\
  }\href@noop {} {\bibfield  {journal} {\bibinfo  {journal} {J. Phys. Chem. C}\
  }\textbf {\bibinfo {volume} {114}},\ \bibinfo {pages} {10962--10968}
  (\bibinfo {year} {2010})}\BibitemShut {NoStop}%
\bibitem [{\citenamefont {Long}\ \emph {et~al.}(2017)\citenamefont {Long},
  \citenamefont {Hedley}, \citenamefont {Ruseckas}, \citenamefont {Chowdhury},
  \citenamefont {Roland}, \citenamefont {Serrano}, \citenamefont {Cooke},\ and\
  \citenamefont {Samuel}}]{long2017}%
  \BibitemOpen
  \bibfield  {author} {\bibinfo {author} {\bibfnamefont {Y.}~\bibnamefont
  {Long}}, \bibinfo {author} {\bibfnamefont {G.~J.}\ \bibnamefont {Hedley}},
  \bibinfo {author} {\bibfnamefont {A.}~\bibnamefont {Ruseckas}}, \bibinfo
  {author} {\bibfnamefont {M.}~\bibnamefont {Chowdhury}}, \bibinfo {author}
  {\bibfnamefont {T.}~\bibnamefont {Roland}}, \bibinfo {author} {\bibfnamefont
  {L.~A.}\ \bibnamefont {Serrano}}, \bibinfo {author} {\bibfnamefont
  {G.}~\bibnamefont {Cooke}}, \ and\ \bibinfo {author} {\bibfnamefont {I.~D.}\
  \bibnamefont {Samuel}},\ }\bibfield  {title} {\enquote {\bibinfo {title}
  {Effect of annealing on exciton diffusion in a high performance small
  molecule organic photovoltaic material},}\ }\href@noop {} {\bibfield
  {journal} {\bibinfo  {journal} {ACS Appl. Mater. Inter.}\ }\textbf {\bibinfo
  {volume} {9}},\ \bibinfo {pages} {14945--14952} (\bibinfo {year}
  {2017})}\BibitemShut {NoStop}%
\bibitem [{\citenamefont {Zhang}\ \emph {et~al.}(2019)\citenamefont {Zhang},
  \citenamefont {Sajjad}, \citenamefont {Blaszczyk}, \citenamefont {Parnell},
  \citenamefont {Ruseckas}, \citenamefont {Serrano}, \citenamefont {Cooke},\
  and\ \citenamefont {Samuel}}]{zhang2019}%
  \BibitemOpen
  \bibfield  {author} {\bibinfo {author} {\bibfnamefont {Y.}~\bibnamefont
  {Zhang}}, \bibinfo {author} {\bibfnamefont {M.~T.}\ \bibnamefont {Sajjad}},
  \bibinfo {author} {\bibfnamefont {O.}~\bibnamefont {Blaszczyk}}, \bibinfo
  {author} {\bibfnamefont {A.~J.}\ \bibnamefont {Parnell}}, \bibinfo {author}
  {\bibfnamefont {A.}~\bibnamefont {Ruseckas}}, \bibinfo {author}
  {\bibfnamefont {L.~A.}\ \bibnamefont {Serrano}}, \bibinfo {author}
  {\bibfnamefont {G.}~\bibnamefont {Cooke}}, \ and\ \bibinfo {author}
  {\bibfnamefont {I.~D.}\ \bibnamefont {Samuel}},\ }\bibfield  {title}
  {\enquote {\bibinfo {title} {Large crystalline domains and an enhanced
  exciton diffusion length enable efficient organic solar cells},}\ }\href@noop
  {} {\bibfield  {journal} {\bibinfo  {journal} {Chem. Mater.}\ }\textbf
  {\bibinfo {volume} {31}},\ \bibinfo {pages} {6548--6557} (\bibinfo {year}
  {2019})}\BibitemShut {NoStop}%
\bibitem [{\citenamefont {Park}\ \emph {et~al.}(2021)\citenamefont {Park},
  \citenamefont {Chandrabose}, \citenamefont {Price}, \citenamefont {Ryu},
  \citenamefont {Lee}, \citenamefont {Shin}, \citenamefont {Wu}, \citenamefont
  {Lee}, \citenamefont {Chen}, \citenamefont {Dai} \emph {et~al.}}]{park2021}%
  \BibitemOpen
  \bibfield  {author} {\bibinfo {author} {\bibfnamefont {S.~Y.}\ \bibnamefont
  {Park}}, \bibinfo {author} {\bibfnamefont {S.}~\bibnamefont {Chandrabose}},
  \bibinfo {author} {\bibfnamefont {M.~B.}\ \bibnamefont {Price}}, \bibinfo
  {author} {\bibfnamefont {H.~S.}\ \bibnamefont {Ryu}}, \bibinfo {author}
  {\bibfnamefont {T.~H.}\ \bibnamefont {Lee}}, \bibinfo {author} {\bibfnamefont
  {Y.~S.}\ \bibnamefont {Shin}}, \bibinfo {author} {\bibfnamefont
  {Z.}~\bibnamefont {Wu}}, \bibinfo {author} {\bibfnamefont {W.}~\bibnamefont
  {Lee}}, \bibinfo {author} {\bibfnamefont {K.}~\bibnamefont {Chen}}, \bibinfo
  {author} {\bibfnamefont {S.}~\bibnamefont {Dai}},  \emph {et~al.},\
  }\bibfield  {title} {\enquote {\bibinfo {title} {Photophysical pathways in
  efficient bilayer organic solar cells: The importance of interlayer energy
  transfer},}\ }\href@noop {} {\bibfield  {journal} {\bibinfo  {journal} {Nano
  Energy}\ }\textbf {\bibinfo {volume} {84}},\ \bibinfo {pages} {105924}
  (\bibinfo {year} {2021})}\BibitemShut {NoStop}%
\bibitem [{\citenamefont {Mikhnenko}\ \emph {et~al.}(2015)\citenamefont
  {Mikhnenko}, \citenamefont {Blom},\ and\ \citenamefont
  {Nguyen}}]{mikhnenko2015}%
  \BibitemOpen
  \bibfield  {author} {\bibinfo {author} {\bibfnamefont {O.~V.}\ \bibnamefont
  {Mikhnenko}}, \bibinfo {author} {\bibfnamefont {P.~W.}\ \bibnamefont {Blom}},
  \ and\ \bibinfo {author} {\bibfnamefont {T.-Q.}\ \bibnamefont {Nguyen}},\
  }\bibfield  {title} {\enquote {\bibinfo {title} {Exciton diffusion in organic
  semiconductors},}\ }\href@noop {} {\bibfield  {journal} {\bibinfo  {journal}
  {Energy Environ. Sci.}\ }\textbf {\bibinfo {volume} {8}},\ \bibinfo {pages}
  {1867--1888} (\bibinfo {year} {2015})}\BibitemShut {NoStop}%
\bibitem [{\citenamefont {Chandrasekhar}(1943)}]{chandrasekhar1943}%
  \BibitemOpen
  \bibfield  {author} {\bibinfo {author} {\bibfnamefont {S.}~\bibnamefont
  {Chandrasekhar}},\ }\bibfield  {title} {\enquote {\bibinfo {title}
  {Stochastic problems in physics and astronomy},}\ }\href@noop {} {\bibfield
  {journal} {\bibinfo  {journal} {Rev. Mod. Phys.}\ }\textbf {\bibinfo {volume}
  {15}},\ \bibinfo {pages} {1} (\bibinfo {year} {1943})}\BibitemShut {NoStop}%
\bibitem [{\citenamefont {Kumar}\ \emph {et~al.}(2014)\citenamefont {Kumar},
  \citenamefont {Cui}, \citenamefont {Ceballos}, \citenamefont {He},
  \citenamefont {Wang},\ and\ \citenamefont {Zhao}}]{kumar2014}%
  \BibitemOpen
  \bibfield  {author} {\bibinfo {author} {\bibfnamefont {N.}~\bibnamefont
  {Kumar}}, \bibinfo {author} {\bibfnamefont {Q.}~\bibnamefont {Cui}}, \bibinfo
  {author} {\bibfnamefont {F.}~\bibnamefont {Ceballos}}, \bibinfo {author}
  {\bibfnamefont {D.}~\bibnamefont {He}}, \bibinfo {author} {\bibfnamefont
  {Y.}~\bibnamefont {Wang}}, \ and\ \bibinfo {author} {\bibfnamefont
  {H.}~\bibnamefont {Zhao}},\ }\bibfield  {title} {\enquote {\bibinfo {title}
  {Exciton-exciton annihilation in {M}o{S}e 2 monolayers},}\ }\href@noop {}
  {\bibfield  {journal} {\bibinfo  {journal} {Phys. Rev. B}\ }\textbf {\bibinfo
  {volume} {89}},\ \bibinfo {pages} {125427} (\bibinfo {year}
  {2014})}\BibitemShut {NoStop}%
\bibitem [{\citenamefont {Linardy}\ \emph {et~al.}(2020)\citenamefont
  {Linardy}, \citenamefont {Yadav}, \citenamefont {Vella}, \citenamefont
  {Verzhbitskiy}, \citenamefont {Watanabe}, \citenamefont {Taniguchi},
  \citenamefont {Pauly}, \citenamefont {Trushin},\ and\ \citenamefont
  {Eda}}]{linardy2020}%
  \BibitemOpen
  \bibfield  {author} {\bibinfo {author} {\bibfnamefont {E.}~\bibnamefont
  {Linardy}}, \bibinfo {author} {\bibfnamefont {D.}~\bibnamefont {Yadav}},
  \bibinfo {author} {\bibfnamefont {D.}~\bibnamefont {Vella}}, \bibinfo
  {author} {\bibfnamefont {I.~A.}\ \bibnamefont {Verzhbitskiy}}, \bibinfo
  {author} {\bibfnamefont {K.}~\bibnamefont {Watanabe}}, \bibinfo {author}
  {\bibfnamefont {T.}~\bibnamefont {Taniguchi}}, \bibinfo {author}
  {\bibfnamefont {F.}~\bibnamefont {Pauly}}, \bibinfo {author} {\bibfnamefont
  {M.}~\bibnamefont {Trushin}}, \ and\ \bibinfo {author} {\bibfnamefont
  {G.}~\bibnamefont {Eda}},\ }\bibfield  {title} {\enquote {\bibinfo {title}
  {Harnessing exciton--exciton annihilation in two-dimensional
  semiconductors},}\ }\href@noop {} {\bibfield  {journal} {\bibinfo  {journal}
  {Nano Lett.}\ }\textbf {\bibinfo {volume} {20}},\ \bibinfo {pages}
  {1647--1653} (\bibinfo {year} {2020})}\BibitemShut {NoStop}%
\bibitem [{\citenamefont {Deshmukh}\ \emph {et~al.}(2018)\citenamefont
  {Deshmukh}, \citenamefont {Matsidik}, \citenamefont {Prasad}, \citenamefont
  {Chandrasekaran}, \citenamefont {Welford}, \citenamefont {Connal},
  \citenamefont {Liu}, \citenamefont {Gann}, \citenamefont {Thomsen},
  \citenamefont {Kabra} \emph {et~al.}}]{deshmukh2018}%
  \BibitemOpen
  \bibfield  {author} {\bibinfo {author} {\bibfnamefont {K.~D.}\ \bibnamefont
  {Deshmukh}}, \bibinfo {author} {\bibfnamefont {R.}~\bibnamefont {Matsidik}},
  \bibinfo {author} {\bibfnamefont {S.~K.}\ \bibnamefont {Prasad}}, \bibinfo
  {author} {\bibfnamefont {N.}~\bibnamefont {Chandrasekaran}}, \bibinfo
  {author} {\bibfnamefont {A.}~\bibnamefont {Welford}}, \bibinfo {author}
  {\bibfnamefont {L.~A.}\ \bibnamefont {Connal}}, \bibinfo {author}
  {\bibfnamefont {A.~C.}\ \bibnamefont {Liu}}, \bibinfo {author} {\bibfnamefont
  {E.}~\bibnamefont {Gann}}, \bibinfo {author} {\bibfnamefont {L.}~\bibnamefont
  {Thomsen}}, \bibinfo {author} {\bibfnamefont {D.}~\bibnamefont {Kabra}},
  \emph {et~al.},\ }\bibfield  {title} {\enquote {\bibinfo {title} {Impact of
  acceptor fluorination on the performance of all-polymer solar cells},}\
  }\href@noop {} {\bibfield  {journal} {\bibinfo  {journal} {ACS Appl. Mater.
  Inter.}\ }\textbf {\bibinfo {volume} {10}},\ \bibinfo {pages} {955--969}
  (\bibinfo {year} {2018})}\BibitemShut {NoStop}%
\bibitem [{\citenamefont {Wong-Stringer}\ \emph {et~al.}(2017)\citenamefont
  {Wong-Stringer}, \citenamefont {Bishop}, \citenamefont {Smith}, \citenamefont
  {Mohamad}, \citenamefont {Parnell}, \citenamefont {Kumar}, \citenamefont
  {Rodenburg},\ and\ \citenamefont {Lidzey}}]{wong2017}%
  \BibitemOpen
  \bibfield  {author} {\bibinfo {author} {\bibfnamefont {M.}~\bibnamefont
  {Wong-Stringer}}, \bibinfo {author} {\bibfnamefont {J.~E.}\ \bibnamefont
  {Bishop}}, \bibinfo {author} {\bibfnamefont {J.~A.}\ \bibnamefont {Smith}},
  \bibinfo {author} {\bibfnamefont {D.~K.}\ \bibnamefont {Mohamad}}, \bibinfo
  {author} {\bibfnamefont {A.~J.}\ \bibnamefont {Parnell}}, \bibinfo {author}
  {\bibfnamefont {V.}~\bibnamefont {Kumar}}, \bibinfo {author} {\bibfnamefont
  {C.}~\bibnamefont {Rodenburg}}, \ and\ \bibinfo {author} {\bibfnamefont
  {D.~G.}\ \bibnamefont {Lidzey}},\ }\bibfield  {title} {\enquote {\bibinfo
  {title} {Efficient perovskite photovoltaic devices using chemically doped
  {PCDTBT} as a hole-transport material},}\ }\href@noop {} {\bibfield
  {journal} {\bibinfo  {journal} {J. Mater. Chem. A}\ }\textbf {\bibinfo
  {volume} {5}},\ \bibinfo {pages} {15714--15723} (\bibinfo {year}
  {2017})}\BibitemShut {NoStop}%
\bibitem [{\citenamefont {Chu}\ \emph {et~al.}(2016)\citenamefont {Chu},
  \citenamefont {Wang}, \citenamefont {Fu}, \citenamefont {Choi}, \citenamefont
  {Park}, \citenamefont {Srinivasarao},\ and\ \citenamefont
  {Reichmanis}}]{chu2016}%
  \BibitemOpen
  \bibfield  {author} {\bibinfo {author} {\bibfnamefont {P.-H.}\ \bibnamefont
  {Chu}}, \bibinfo {author} {\bibfnamefont {G.}~\bibnamefont {Wang}}, \bibinfo
  {author} {\bibfnamefont {B.}~\bibnamefont {Fu}}, \bibinfo {author}
  {\bibfnamefont {D.}~\bibnamefont {Choi}}, \bibinfo {author} {\bibfnamefont
  {J.~O.}\ \bibnamefont {Park}}, \bibinfo {author} {\bibfnamefont
  {M.}~\bibnamefont {Srinivasarao}}, \ and\ \bibinfo {author} {\bibfnamefont
  {E.}~\bibnamefont {Reichmanis}},\ }\bibfield  {title} {\enquote {\bibinfo
  {title} {Synergistic effect of regioregular and regiorandom poly
  (3-hexylthiophene) blends for high performance flexible organic field effect
  transistors},}\ }\href@noop {} {\bibfield  {journal} {\bibinfo  {journal}
  {Adv. Electron. Mater.}\ }\textbf {\bibinfo {volume} {2}},\ \bibinfo {pages}
  {1500384} (\bibinfo {year} {2016})}\BibitemShut {NoStop}%
\bibitem [{\citenamefont {Mai}\ \emph {et~al.}(2018)\citenamefont {Mai},
  \citenamefont {Xiao}, \citenamefont {Zhou}, \citenamefont {Wang},
  \citenamefont {Zhu}, \citenamefont {Zhao}, \citenamefont {Zhan},\ and\
  \citenamefont {Lu}}]{mai2018}%
  \BibitemOpen
  \bibfield  {author} {\bibinfo {author} {\bibfnamefont {J.}~\bibnamefont
  {Mai}}, \bibinfo {author} {\bibfnamefont {Y.}~\bibnamefont {Xiao}}, \bibinfo
  {author} {\bibfnamefont {G.}~\bibnamefont {Zhou}}, \bibinfo {author}
  {\bibfnamefont {J.}~\bibnamefont {Wang}}, \bibinfo {author} {\bibfnamefont
  {J.}~\bibnamefont {Zhu}}, \bibinfo {author} {\bibfnamefont {N.}~\bibnamefont
  {Zhao}}, \bibinfo {author} {\bibfnamefont {X.}~\bibnamefont {Zhan}}, \ and\
  \bibinfo {author} {\bibfnamefont {X.}~\bibnamefont {Lu}},\ }\bibfield
  {title} {\enquote {\bibinfo {title} {Hidden structure ordering along backbone
  of fused-ring electron acceptors enhanced by ternary bulk heterojunction},}\
  }\href@noop {} {\bibfield  {journal} {\bibinfo  {journal} {Adv. Mater.}\
  }\textbf {\bibinfo {volume} {30}},\ \bibinfo {pages} {1802888} (\bibinfo
  {year} {2018})}\BibitemShut {NoStop}%
\bibitem [{\citenamefont {Ma}\ \emph {et~al.}(2020)\citenamefont {Ma},
  \citenamefont {Chen}, \citenamefont {Liu}, \citenamefont {Xiao},
  \citenamefont {Luo}, \citenamefont {Zhang}, \citenamefont {Luo},
  \citenamefont {Lu}, \citenamefont {Zhang}, \citenamefont {Li} \emph
  {et~al.}}]{ma2020}%
  \BibitemOpen
  \bibfield  {author} {\bibinfo {author} {\bibfnamefont {R.}~\bibnamefont
  {Ma}}, \bibinfo {author} {\bibfnamefont {Y.}~\bibnamefont {Chen}}, \bibinfo
  {author} {\bibfnamefont {T.}~\bibnamefont {Liu}}, \bibinfo {author}
  {\bibfnamefont {Y.}~\bibnamefont {Xiao}}, \bibinfo {author} {\bibfnamefont
  {Z.}~\bibnamefont {Luo}}, \bibinfo {author} {\bibfnamefont {M.}~\bibnamefont
  {Zhang}}, \bibinfo {author} {\bibfnamefont {S.}~\bibnamefont {Luo}}, \bibinfo
  {author} {\bibfnamefont {X.}~\bibnamefont {Lu}}, \bibinfo {author}
  {\bibfnamefont {G.}~\bibnamefont {Zhang}}, \bibinfo {author} {\bibfnamefont
  {Y.}~\bibnamefont {Li}},  \emph {et~al.},\ }\bibfield  {title} {\enquote
  {\bibinfo {title} {Improving the performance of near infrared binary polymer
  solar cells by adding a second non-fullerene intermediate band-gap
  acceptor},}\ }\href@noop {} {\bibfield  {journal} {\bibinfo  {journal} {J.
  Mater. Chem. C}\ }\textbf {\bibinfo {volume} {8}},\ \bibinfo {pages}
  {909--915} (\bibinfo {year} {2020})}\BibitemShut {NoStop}%
\bibitem [{\citenamefont {Kaiser}\ \emph {et~al.}(2018)\citenamefont {Kaiser},
  \citenamefont {Popp}, \citenamefont {Rinderle}, \citenamefont {Albes},\ and\
  \citenamefont {Gagliardi}}]{kaiser2018}%
  \BibitemOpen
  \bibfield  {author} {\bibinfo {author} {\bibfnamefont {W.}~\bibnamefont
  {Kaiser}}, \bibinfo {author} {\bibfnamefont {J.}~\bibnamefont {Popp}},
  \bibinfo {author} {\bibfnamefont {M.}~\bibnamefont {Rinderle}}, \bibinfo
  {author} {\bibfnamefont {T.}~\bibnamefont {Albes}}, \ and\ \bibinfo {author}
  {\bibfnamefont {A.}~\bibnamefont {Gagliardi}},\ }\bibfield  {title} {\enquote
  {\bibinfo {title} {Generalized kinetic {M}onte--{C}arlo framework for organic
  electronics},}\ }\href@noop {} {\bibfield  {journal} {\bibinfo  {journal}
  {Algorithms}\ }\textbf {\bibinfo {volume} {11}},\ \bibinfo {pages} {37}
  (\bibinfo {year} {2018})}\BibitemShut {NoStop}%
\bibitem [{\citenamefont {Gillespie}(1977)}]{gillespie}%
  \BibitemOpen
  \bibfield  {author} {\bibinfo {author} {\bibfnamefont {D.~T.}\ \bibnamefont
  {Gillespie}},\ }\bibfield  {title} {\enquote {\bibinfo {title} {Exact
  stochastic simulation of coupled chemical reactions},}\ }\href@noop {}
  {\bibfield  {journal} {\bibinfo  {journal} {J. Phys. Chem.}\ }\textbf
  {\bibinfo {volume} {81}},\ \bibinfo {pages} {2340--2361} (\bibinfo {year}
  {1977})}\BibitemShut {NoStop}%
\bibitem [{\citenamefont {Gerischer}\ and\ \citenamefont
  {Heller}(1991)}]{gerischer1991}%
  \BibitemOpen
  \bibfield  {author} {\bibinfo {author} {\bibfnamefont {H.}~\bibnamefont
  {Gerischer}}\ and\ \bibinfo {author} {\bibfnamefont {A.}~\bibnamefont
  {Heller}},\ }\bibfield  {title} {\enquote {\bibinfo {title} {The role of
  oxygen in photooxidation of organic molecules on semiconductor particles},}\
  }\href@noop {} {\bibfield  {journal} {\bibinfo  {journal} {J. Phys. Chem.}\
  }\textbf {\bibinfo {volume} {95}},\ \bibinfo {pages} {5261--5267} (\bibinfo
  {year} {1991})}\BibitemShut {NoStop}%
\bibitem [{\citenamefont {Yoshida}\ and\ \citenamefont
  {Asakura}(1976)}]{yoshida1976}%
  \BibitemOpen
  \bibfield  {author} {\bibinfo {author} {\bibfnamefont {A.}~\bibnamefont
  {Yoshida}}\ and\ \bibinfo {author} {\bibfnamefont {T.}~\bibnamefont
  {Asakura}},\ }\bibfield  {title} {\enquote {\bibinfo {title} {A simple
  technique for quickly measuring the spot size of gaussian laser beams},}\
  }\href@noop {} {\bibfield  {journal} {\bibinfo  {journal} {Opt. Laser
  Technol.}\ }\textbf {\bibinfo {volume} {8}},\ \bibinfo {pages} {273--274}
  (\bibinfo {year} {1976})}\BibitemShut {NoStop}%
\end{thebibliography}%

\end{document}